\documentclass[journal=acsmacrolett,manuscript=letter,layout=twocolumn]{achemso}

\setkeys{acs}{articletitle = true} 

\usepackage{array}
\usepackage{natbib}
\usepackage{stfloats}
\usepackage{chemformula} 
\usepackage[T1]{fontenc} 
\usepackage{float} %
\usepackage{seqsplit} 
\usepackage{graphicx}
\usepackage{booktabs} 
\usepackage{multirow}
\usepackage{color, colortbl}
\usepackage[para,flushleft]{threeparttable}
\usepackage{ulem}
\usepackage{graphicx}
\usepackage[super]{nth}
\usepackage{amsmath}
\usepackage[font=scriptsize,labelfont=bf]{caption}
\usepackage[switch]{lineno}
\usepackage{breqn}
\usepackage[version=4]{mhchem}
\usepackage{adjustbox}
\usepackage{makecell} 
\DeclareUnicodeCharacter{2009}{\,} 
\usepackage[utf8]{inputenc}
\DeclareUnicodeCharacter{00FC}{\"u}
\mciteErrorOnUnknownfalse

\author{Binay P. Nayak}
\affiliation{Ames National Laboratory, and Department of Chemical and Biological Engineering, Iowa State University, Ames, Iowa 50011, United States}

\author{Hyeong Jin Kim}
\affiliation{Ames National Laboratory, and Department of Chemical and Biological Engineering, Iowa State University, Ames, Iowa 50011, United States}

\author{Srikanth Nayak}
\affiliation{Ames National Laboratory, and Department of Chemical and Biological Engineering, Iowa State University, Ames, Iowa 50011, United States}
\altaffiliation{Current address: Chemical Sciences and Engineering Division, Argonne National Laboratory, Lemont, Illinois 60439, United States}

\author{Wenjie Wang}
\affiliation{Division of Materials Sciences and Engineering, Ames National Laboratory, U.S. DOE, Ames, Iowa 50011, United States}

\author{Wei Bu}
\affiliation{NSF’s ChemMatCARS, Pritzker School of Molecular Engineering, University of Chicago, Chicago, Illinois 60637, United States}

\author{Surya K. Mallapragada}
\email{suryakm@iastate.edu}
\affiliation{Ames National Laboratory, and Department of Chemical and Biological Engineering, Iowa State University, Ames, Iowa 50011, United States}

\author{David Vaknin}
\affiliation{Ames National Laboratory, and Department of Physics and Astronomy, Iowa State University, Ames, Iowa 50011, United States}
\email{vaknin@ameslab.gov}

\title
{Assembling PNIPAM-Capped Gold Nanoparticles in Aqueous Solutions}

\begin{document}

\begin{abstract}
Employing small angle X-ray scattering (SAXS), we explore the conditions under which assembly of gold nanoparticles (AuNPs) grafted with the thermo-sensitive polymer Poly(\textit{N}-isopropylacrylamide) (PNIPAM) emerges. We find that short-range order assembly emerges by combining the addition of electrolytes or poly-electrolytes with raising the temperature of the suspensions above the lower-critical solution temperature (LCST) of PNIPAM. Our results show that the longer the PNIPAM chain is, the better organization in the assembled clusters. Interestingly, without added electrolytes, there is no evidence of AuNPs assembly as a function of temperature, although untethered PNIPAM is known to undergo a coil-to-globule transition above its LCST. This study demonstrates another approach to assembling potential thermo-sensitive nanostructures for devices by leveraging the unique properties of PNIPAM.

\end{abstract}

\section{Main}

Poly(\textit{N}-isopropylacrylamide) (PNIPAM) is an amphiphilic polymer comprising an alkyl-chain back-bone decorated with amide-isopropyl side groups. The amide side groups, common to protein chains, render hydrophilic properties to the polymer \textbf{}. PNIPAM has attracted attention across disciplines due to its unique thermally responsive behavior. The polymer exhibits a lower critical solution temperature (LCST) at $\sim$ 32 $^ \circ$C,\cite{zhulina1991coil,heskins1968solution} above which the chains expel water and undergo contraction to a cascade of globular conformations.\cite{wu1998globule} It has been established that the LCST phase transition is reversible. In addition, small angle neutron scattering of PNIPAM suspensions shows evidence of reversible assembly of the globular structures.\cite{filippov2016internal} This unique property has been widely explored for drug delivery\cite{xiong2011dual,cao2019reversible},bio-sensors\cite{ma2022nanochitin,guan2011pnipam}, smart layers \cite{wang2009cyclic,rotzetter2012thermoresponsive}, and microactuator \cite{zhang2011optically,tu2017self}. The thermal properties of PNIPAM make it a suitable candidate for surface modifications of nanoparticles (NPs) to create stimuli-responsive self-assembly and crystallization. \cite{ding2016light,cormier2018dynamic}

Recently, PNIPAM has been synthesized with a thiol end-group, making it suitable for grafting metallic NPs, particularly gold and silver. \cite{li2019synthesis} Indeed, temperature-induced assembly PNIPAM grafted nanoparticles have been observed above the LCST by varying salinity, pH, and by photoexcitation \cite{gibson2013aggregate,zhang2013salt,jones2016importance,maji2016poly,li2014poly}. Various dynamic light scattering (DLS) and ultraviolet-visible (UV-Vis) studies have shown that the hydrodynamic diameter ($D_{\rm H}$) of PNIPAM grafted AuNPs in pure water decreases marginally above the LCST. However, upon adding sodium chloride to the solutions, aggregation emerges above the LCST.\cite{yusa2007salt,vasicek2017thermoresponsive,turek2018crucial,maji2016poly,zhang2013salt,gibson2013aggregate}   Using block copolymer, poly(ethylene glycol)-b-poly(\textit{N}-isopropylacrylamide) to graft AuNPs, it has been shown that self-assembly into one-dimensional (1D) or two-dimensional (2D) structures in salt solutions can be induced by raising the temperature above the LCST.\cite{kim2023directional} The same study emphasizes the significance of adding charged molecules to the grafted NP suspensions to achieve assembly. Other studies of grafted AuNPs with PNIPAM have been shown to exhibit assembly in two dimensions at air/liquid interfaces.\cite{wang2018two,minier2021poly,londono2021salt} Although thermal effects have not been reported to achieve assembly, the polymer tends to respond to the salinity of the suspension in a similar manner as has been observed for poly-ethylene glycol (PEG) grafted AuNPs \cite{zhang2017macroscopic,kim2021nanoparticleNTE}. 

Here, we extend these 2D studies to the three-dimensional (3D) bulk self-assembly and ordering by monitoring the combined effect of salinity and temperature. DLS studies have indicated assembly upon a variable salinity and temperature combination. \cite{kim2023directional} We employ \textit{in-situ} synchrotron-based small angle scattering (SAXS) technique to determine the nature of the assembly upon adding electrolytes and varying the temperature.\cite{Nayak2019IPC} As for electrolytes, we use salts such as potassium carbonate (\ch{K2CO3}), sodium chloride (NaCl), or long-chain positively charged ploy-electrolyte Poly(diallyl-dimethylammonium chloride) (PDAC). PDAC has been shown to induce 2D crystallization of sodium do-decyl sulfide at the air/liquid interface, making it a potential electrolyte to facilitate assembly.\cite{vaknin2004induced} We also examine the effect of grafted PNIPAM molecular weight ($\sim$ 3 vs. 6 kDa) on the characteristics of the assembly.

\begin{figure}[!ht]
 	\centering 
 	\includegraphics[width=\linewidth]{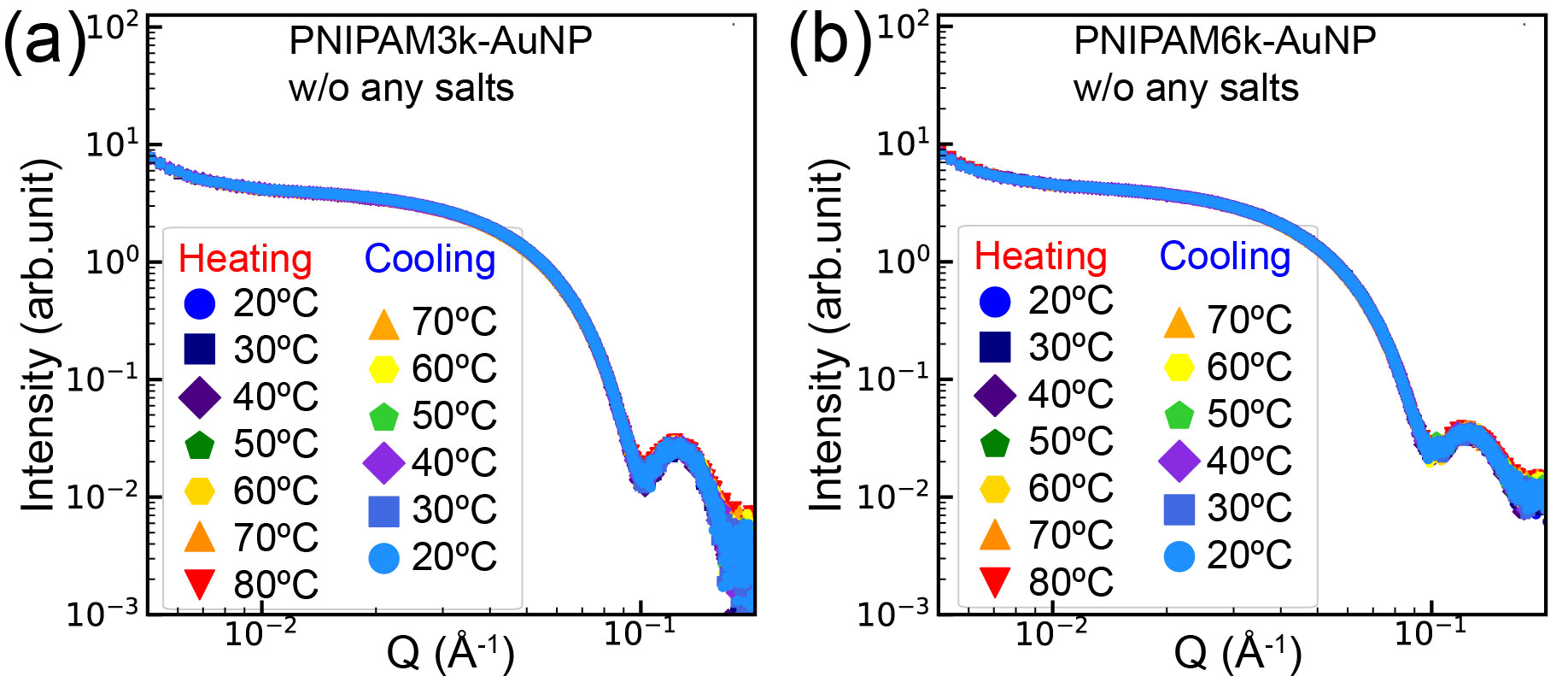}
    \caption{ SAXS data for (a) PNIPAM3k-AuNPs (10 nm core) and (b) PNIPAM6k-AuNPs (10 nm core) in water (i.e., without any electrolytes) at various temperatures as indicated. The normalized intensities profiles $S(Q)$ are shown in SI Figure \ref{form_temp_s} and prove that no assembly occurs upon raising the temperature.
 	}
 	\label{form_temp_i}
 \end{figure} 
 
Raising the temperature above the LCST, without adding any electrolytes, the PNIPAM-AuNPs remain dispersed in the suspensions.  
Figure \ref{form_temp_i} shows SAXS patterns obtained from PNIPAM-AuNPs without salts at various temperatures (heating and cooling cycles). The pattern for PNIPAM3k-AuNPs in (a) and PNIPAM6k-AuNPs (b), up to 80 \textdegree{}C consists of the form factor of the core AuNPs (See Figure \ref{form_fit}). We note that the SAXS intensities are dominated by the form factor of the AuNP core with little contribution from the PNIPAM corona. These results do not provide clear evidence for conformational change above the LCST of PNIPAM-AuNPs. We conclude that the particles remain dispersed in the suspensions even above the LCST. This is consistent with the globular shrinking conformation above the LCST,\cite{wu1998globule} where the polymer likely exposes its hydrophilic moieties to the aqueous medium. The absence of assembly in pure water above LCST can be rationalized by the repulsion between the hydrophilic (dipolar) moieties. \cite{zhang2013salt,turek2018crucial} 
The lack of scattering from the PNIPAM corona in aqueous suspensions is due to a negligible electron-density (ED) contrast between the suspension (water) and the organic polymer. As a result, one cannot infer from SAXS measurements moderate changes in the conformations of the polymer in the corona. The normalized intensity profiles $S(Q)$ are shown in Figure \ref{form_temp_s} of the SI, confirming well-dispersed grafted AuNPs at all measured temperatures. More details on the form factor, the core size of AuNPs, and the size distribution of the citrate-stabilized AuNPs are provided in the SI.

\begin{figure*}[!ht]
 	\centering 
 	\includegraphics[width=0.7 \linewidth]{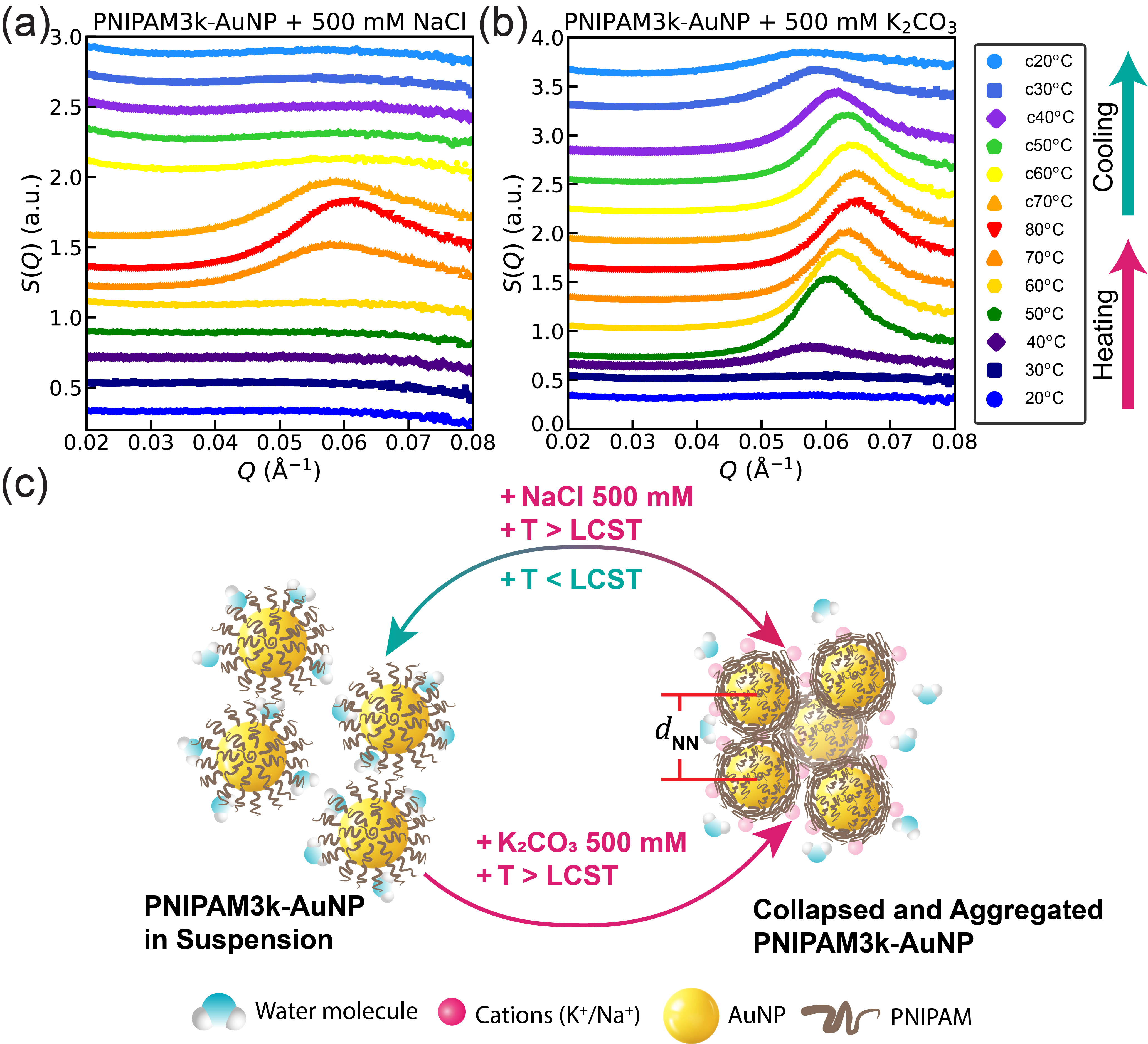}
 	\caption{Normalized intensity $S(Q)$ data for PNIPAM3k-AuNPs with (a) 500 mM NaCl, (b) 500 mM \ch{K2CO3} at various temperatures, as indicated, showing the emergence of broad interference peak at $Q \Large \simeq 0.055$ \AA$^{-1}$. Such a lone broad peak indicates amorphous aggregation of particles with a characteristic nearest neighbor distance $d_{\rm NN}\sim$ 11.5 nm. (c) Schematic illustration of the transition from dispersed nanoparticles to aggregates as the temperature is raised above the LCST in the presence of salts. The depicted aggregates show the particles with collapsed PNIPAM corona inferred from the value of $d_{\rm NN}$, which is close to the NP core diameter.}
 	\label{3d_3k_main}
 \end{figure*}  

\begin{figure*}[!hbt]
 	\centering 
 	\includegraphics[width=0.7 \linewidth]{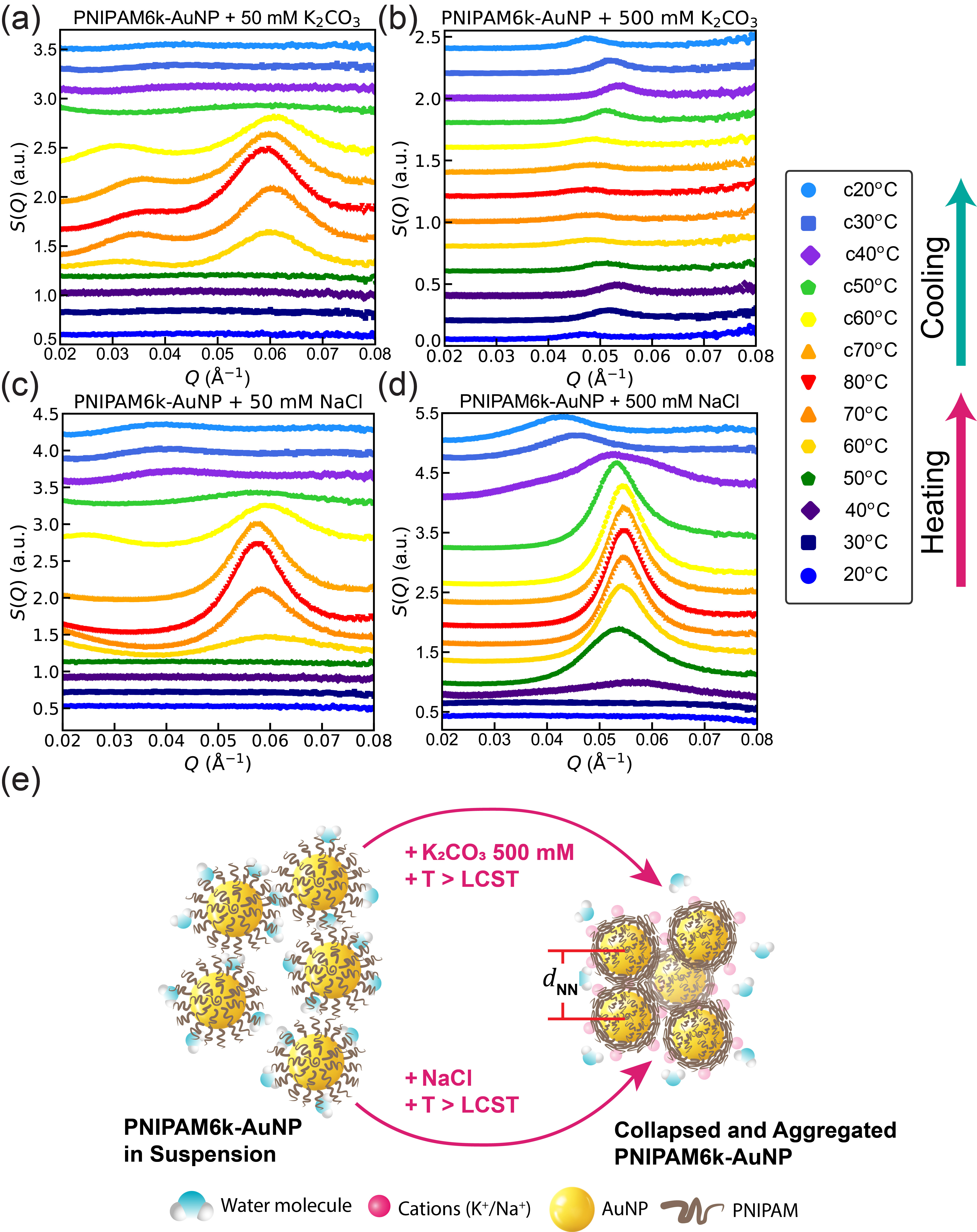}
 	\caption{Normalized intensity $S(Q)$ patterns for PNIPAM6k-AuNPs with (a) 50 mM \ch{K2CO3}, showing two broad diffraction peaks. In the SI (Figure \ref{diamond}), we show that the diffraction pattern is best-describing clusters with diamond-like motifs. Our conclusion is based on examining various other structural scenarios. (b) 500 mM \ch{K2CO3} shows a single weak peak corresponding to a $d_{\rm NN} \simeq 12.5$ nm. It is also likely that at this concentration, precipitation occurs. (c) 50 mM NaCl, (d) 500 mM NaCl at various temperatures as indicated, showing a single peak as described in Figure \ref{3d_3k_main}. (e) Schematic illustrations of the assembly development from dispersed NPs to aggregated are presented. Note that in this case, the assembly process is irreversible.}
 	\label{3d_6k_main} 
 \end{figure*}
 
Raising the temperature above the LCST in the presence of salts induces aggregation of PNIAPM3k-AuNPs. At low salt concentrations (below 50 mM of NaCl or \ch{K2CO3}), the SAXS data show that the particles are dispersed in the suspensions even at elevated temperatures above the LCST, as shown in Figure \ref{3k_si}. Figure \ref{3d_3k_main} shows SAXS $S(Q)$ patterns of  PNIPAM3k-AuNPs at 500 mM (a) NaCl and (b) \ch{K2CO3}. Adding \ch{K2CO3} or NaCl to the solution at room temperature yields SAXS patterns that are similar to those shown in Figure \ref{form_temp_i} (i.e., without salts). However, heating the same salinated suspensions above $\sim \mathrm{40}~ ^\circ$C gives rise to the emergence of a prominent peak around $Q_{0} = 0.06 $ \AA$^{-1}$,  which gradually shifts to a higher $Q$ value upon further increase in temperature. This stand-alone peak indicates random aggregation of NPs with a characteristic nearest-neighbor (NN) distance ($d_{\rm NN}$), indicating liquid-like order. The characteristic $d_{\rm NN} \simeq \frac{2\pi}{Q_{0}} \simeq 10.4$ nm is slightly larger than the diameter of the core AuNPs ($D_{\rm Core}$ = 8.7 nm, see Figure \ref{form_fit}). Such a small $d_{\rm NN}$ close to the $D_{\rm Core}$ suggests that the polymer is likely collapsed to its densely packed state (void of water) onto the NP surface. In the SI, we determine an upper limit to the grafting density, assuming such a densely packed collapsed corona. Furthermore, the shift in peak position to higher $Q$ values as the temperature increases indicates smaller $d_{\rm NN}$ and further collapse of the grafted PNIPAM corona consistent with the globular behavior of pure PNIPAM in aqueous solutions.\cite{wu1998globule} We define collapse as a densely-packed dry polymer with its hydrophilic moieties exposed to the aqueous medium.\cite{minier2021poly} We hypothesize that the observed aggregation is induced by the presence of the cations and anions that lead to attractive interaction among the NPs. We argue that the ions decorate different parts of the polymer corona, leading to weak mutual binding. As a result, upon cooling the suspension from 80 \textdegree{}C to room temperature, the NPs seem to re-disperse in the suspensions, as the hydrophilic moieties are less exposed. The evidence for re-dispersed clusters is that the prominent peak at $Q_{0}$ broadens significantly and almost diminishes upon cooling, indicative of the reversible nature of the collapsed state. We note that the addition of \ch{K2CO3} leads to a sharper $S(Q)$ peak and at larger $Q$ values compared to those obtained by adding NaCl. This indicates that \ch{K2CO3}, at elevated temperatures, leads to higher densely packed polymer corona with more well-defined $d_{\rm NN}$. Our analysis of the diffraction patterns yields peak positions and line widths as shown in Table \ref{tbl:lattice}. The line widths indicate that the correlation lengths in the ordered states for 500 mM \ch{K2CO3} and NaCl is on the order of $\sim$90 and 60 nm, respectively (i.e., 8-5 correlated NN). In addition, the aggregation is not fully reversible in the presence of \ch{K2CO3}. We note that \ch{K2CO3}, unlike NaCl, releases a divalent anion, i.e., \ch{CO3$^{–2}$} whereas NaCl has a monovalent anion. More importantly, \ch{K2CO3} affects the pH (increases the alkalinity) of the suspension. We hypothesize that these differences affect the behavior of the assembled particles.  In fact, assembling PEG-grafted AuNPs shows notable assembly differences between NaCl and \ch{K2CO3} in addition to the suspensions.\cite{zhang2017ion}

\begin{figure}[!ht]
 	\centering 
 	\includegraphics[width=\linewidth]{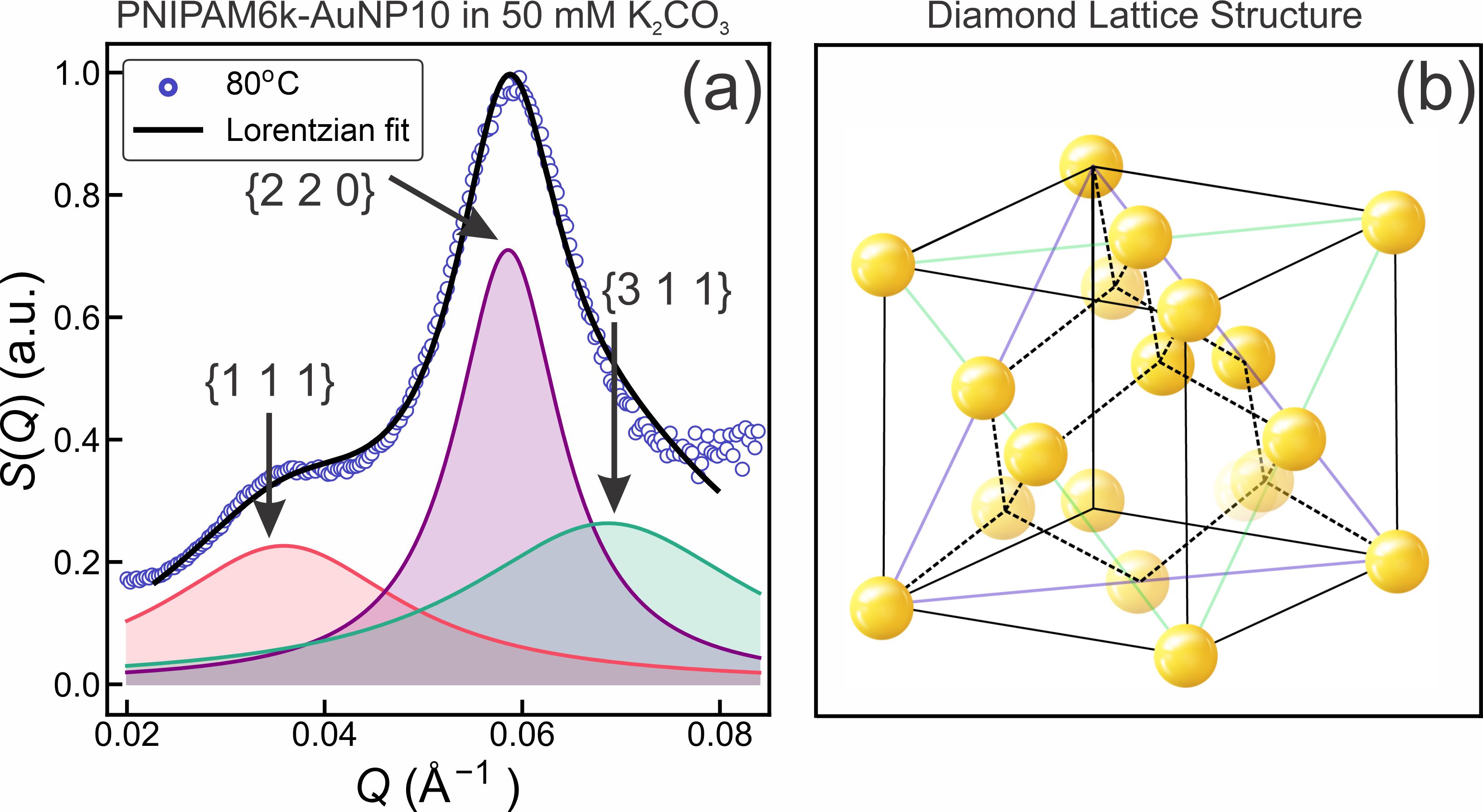}
    \caption{(a) $S(Q)$ profile of PNIPAM6k-AuNP10 at 50 mM \ch{K2CO3} and 80 \textdegree{}C, fitted to a relaxed diamond-like structure using the first three Bragg reflection peaks. The model system accounts for lattice positions with which we calculate the structure factor. The shaded peaks show the contribution of each Bragg reflection to the fitted data. (b) Schematic of a diamond-cubic lattice where the golden spheres represent the collapsed PNIPAM-AuNPs. 
 	}
 	\label{diamond_fit}
 \end{figure}

Similar to PNIPAM3k-AuNPs, the addition of NaCl or \ch{K2CO3} to the suspensions of PNIPAM6k-AuNP has little effect below the LCST, even at concentrations of salt as high as 500 mM. A more noticeable effect of salt addition with temperature for PNIPAM6k-AuNPs is apparent at 50 mM \ch{K2CO3} above the LCST. As shown in Figure \ref{3d_6k_main}(a), two broad peaks ($Q_1 \sim 0.035$ and $Q_2 \sim 0.06$ \AA$^{-1}$) appear upon heating the suspension above 50 \textdegree{}C. Although it is difficult to assign a definite structure from such a limited diffraction pattern, we rationalize our proposed structure based on the behavior of the polymer at different temperatures. In particular, we assume that above the LCST, the polymer corona collapses onto the core of the AuNP. This constraint limits the possible packings of assembled nanoparticles. In the SI, we examine various structural scenarios and conclude that the likely packing has diamond-like motifs,\cite{kalsin2006electrostatic}albeit at very short-range order. The correlation length in the ordered states is of the order of 2-3 unit cells. Figure \ref{diamond_fit}(a) shows $S(Q)$ profile of PNIPAM6k-AuNP10 at 50 mM \ch{K2CO3} and 80 \textdegree{}C, fitted to a relaxed diamond-like structure using the first three Bragg reflection peaks. Our model allows small variations in the lattice positions of the AuNPs. The model system accounts for lattice positions with which we calculate the structure factor. A similar diffraction pattern associated with diamond-like structures has been reported for assembled binary Au and Ag NP systems. \cite{kalsin2006electrostatic}
This interpretation yields a $d_{\rm NN} (\frac{3\pi}{2Q_{111}}) \simeq 13.1$ nm, which is consistent with the grafting density and the fact that the polymer is densely packed (collapsed). See Figure \ref{diamond} and further discussions in the SI. 

At a high concentration of \ch{K2CO3} (500 mM), the diffraction pattern consists of a single weak peak at $Q \sim 0.05$ \AA$^{-1}$ as shown in Figure \ref{3d_6k_main}(b). As mentioned above, a single peak can only provide minimal information on the characteristic length scale of $d_{\rm NN}$; in this case, $d_{\rm NN} \sim 12.5 $ nm. It is also possible that better quality crystals are formed and precipitate out of the suspension and, therefore, are not detected in our bulk solution SAXS measurements. In the SI, we show the assembly of PNIPAM6k-AuNP at the liquid/vapor interface in the presence of 100 mM \ch{K2CO3}. The 2D diffraction pattern (grazing-incidence small-angle X-ray scattering; GISAXS) in Figure \ref{2d_3k6k_main} shows two broad diffraction peaks similar to those observed in the bulk SAXS, however, at slightly smaller $Q$ values. This suggests that the packing at the liquid/vapor interface is similar to that in bulk, and the $d_{\rm NN}$ is slightly larger at the surface. This is expected as the GISAXS measurements are performed below the LCST. We also note that the threshold for ordering in 2D is less strict than in 3D. In 2D, it is sufficient to achieve surface assembly with increased salinity even below the LCST at room temperature.  By contrast, 3D assembly is induced by the combination of salinity and elevated temperature. More evidence on the 2D assembly of PNIPAM6k-AuNPs is established with X-ray reflectivity measurements and their analysis as shown in Figure \ref{2d_3k6k_main}. 

The addition of 50 and 500 mM NaCl to the PNIPAM6k-AuNPs suspension has a similar effect on the assembly. As shown in Figure \ref{3d_6k_main} (c) and (d), the addition of salt has little effect below the LCST. Upon heating, a single peak at about $Q \sim 0.055$ \AA$^{-1}$ emerges and grows in intensity up to 80 \textdegree{}C and decreases in intensity and shifts to lower $Q$ values (at 20 \textdegree{}C the peak is centered around $Q \sim 0.045$ \AA$^{-1}$) demonstrating the reversibility of collapsed state. The observed peak above the LCST is associated with a $d_{\rm NN} \sim 11.5 $ nm and below the LCST $d_{\rm NN} \sim 14.1 $ nm consistent with the globular shrinking of the PNIPAM above the LCST and expanding below it. For NaCl at 50 and 500 mM, we find that the Correlation length is in order of 50 and 100 nm, respectively.

To generalize the effect of assembly using polyelectrolytes, we use PDAC as an additive to the suspension to induce assembly. Figure \ref{PNIPAM_PDAC} (a) shows the normalized intensity vs $Q$ for the grafted AuNPs in the presence of $\sim$ 1 wt $\%$ PDAC. Unlike the simple salts, the polyelectrolyte induces assembly below the LCST (See Figure \ref{PNIPAM_PDAC}). At room temperature, the SAXS pattern has a broad peak at about $Q_{0} = 0.04$ \AA $^{-1}$ which, upon raising the temperature above 35 \textdegree{}C, shifts to $Q_{0} = 0.053$ \AA $^{-1}$. As discussed above, the single peak indicates $d_{\rm NN} = 11.8 $ nm, consistent with a collapsed PNIPAM corona. More details are provided in the SI. 

In summary, we have successfully grafted AuNPs with PNIPAM to achieve temperature-induced assembly and ordering of the NPs. Using synchrotron-based SAXS, we find that temperature has little effect on the nanoparticle assembly in the absence of salts. In fact, the SAXS provides clear evidence that in the absence of salts, the grafted AuNPs are well dispersed in the suspension, even upon heating above LCST. This may be due to the fact that above the LCST, PNIPAM exposes its hydrophilic moieties in the aqueous medium and becomes more soluble. By adding electrolytes (such as \ch{K2CO3}, NaCl, or long chain poly-electrolyte PDAC) to the solution, aggregation emerges. We hypothesize that the mutual attractive interaction among NPs is due to the accumulation of cations and anions on the surfaces of the polymer corona. These interactions lead only to very short-range order assembly such that the SAXS diffraction patterns resemble those of liquids. Our results suggest that the longer the PNIPAM chain, the better organization in the assembled clusters.

\section{Supporting Information}
The Supporting Information is available free of charge on the ACS Publications website at DOI: xxxxx/yyyyy.

Experimental section; Additional DLS data; Additional SAXS data; Structural analysis; 2D XRR and GISAXS data; Calculation of grafting density; Calculation for molarity

\section{Acknowledgements}
The authors thank Jack Lawrence for help in grafting PNIPAM to gold nanoparticles. The authors acknowledge the infrastructure and support provided by the staff at beamline 12-ID-B, Advanced Photon Source (APS), Argonne National Laboratory. The research was financially supported by the U.S. Department of Energy (U.S. DOE), Office of Basic Energy Sciences, Division of Materials Sciences and Engineering. Iowa State University operates Ames National Laboratory for the U.S. DOE under Contract DE-AC02-07CH11358. Part of this research used NSF’s ChemMatCARS Sector 15. NSF’s ChemMatCARS Sector 15 is supported by the Divisions of Chemistry (CHE) and Materials Research (DMR), National Science Foundation, under grant number NSF/CHE-1834750. The use of the Advanced Photon Source, an Office of Science User Facility operated for the U.S. Department of Energy (DOE) Office of Science by Argonne National Laboratory, was supported by the U.S. DOE under Contract No. DE-AC02-06CH11357.  

\section{Author contributions}
WW, DV, and SM conceived and supervised the project. HK, SN, DV, and WW designed, conducted the experiments and analyzed the data. BN, HK, WW, and DV wrote the manuscript. WB supported in X-ray scattering experiments, data acquisition, and data processing at NSF’s ChemMatCARS. SM, DV, and WW secured the funding for the project.  All co-authors read and reviewed the manuscript. 

\normalem
\bibliography{Reference.bib}
\clearpage
\onecolumn

\clearpage
\onecolumn

\setcounter{page}{1}
\setcounter{figure}{0}
\setcounter{equation}{0}
\setcounter{table}{0}

\renewcommand{\thefigure}{TOC\arabic{figure}}
\renewcommand{\theequation}{TOC\arabic{equation}}
\renewcommand{\thetable}{TOC\arabic{table}}
\renewcommand{\thepage}{TOC\arabic{page}} 

\section{{\Large For Table of Contents Use Only}}
\begin{center} 
	{\textbf{\LARGE{Assembling PNIPAM-Capped Gold Nanoparticles in Aqueous Solutions}}}\\
	\bigskip
	\normalsize
    Binay P. Nayak,$^\dagger$ 
    Hyeong Jin Kim,$^\dagger$
    Srikanth Nayak,$^{\dagger, \mid \mid}$
    Wenjie Wang, $^\ddagger$
    Wei Bu,$^\mathparagraph$ 
    Surya K. Mallapragada,$^*$$^,$$^\dagger$ and David Vaknin$^*$$^,$$^\S$\\
	\bigskip
	{$\dagger$\it Ames National Laboratory, and Department of Chemical and Biological Engineering, Iowa State University, Ames, Iowa 50011, United States}\\
	{$\ddagger$\it Division of Materials Sciences and Engineering, Ames National Laboratory, U.S. DOE, Ames, Iowa 50011, United States}\\
	{$\mathparagraph$\it NSF’s ChemMatCARS, Pritzker School of Molecular Engineering, University of Chicago, Chicago, Illinois 60637, United States}\\
	{$\S$\it Ames National Laboratory, and Department of Physics and Astronomy, Iowa State University, Ames, Iowa 50011, United States}\\
    {$\mid \mid$\it Current address: Chemical Sciences and Engineering Division, Argonne National Laboratory, Lemont, Illinois 60439, United States}\\
 
	\bigskip
	{E-mail: suryakm@iastate.edu; vaknin@ameslab.gov}\\
\end{center}

\subsection{Table of Contents Graphic}
\begin{figure}[H]
    \centering
    \includegraphics[width=0.9\linewidth] {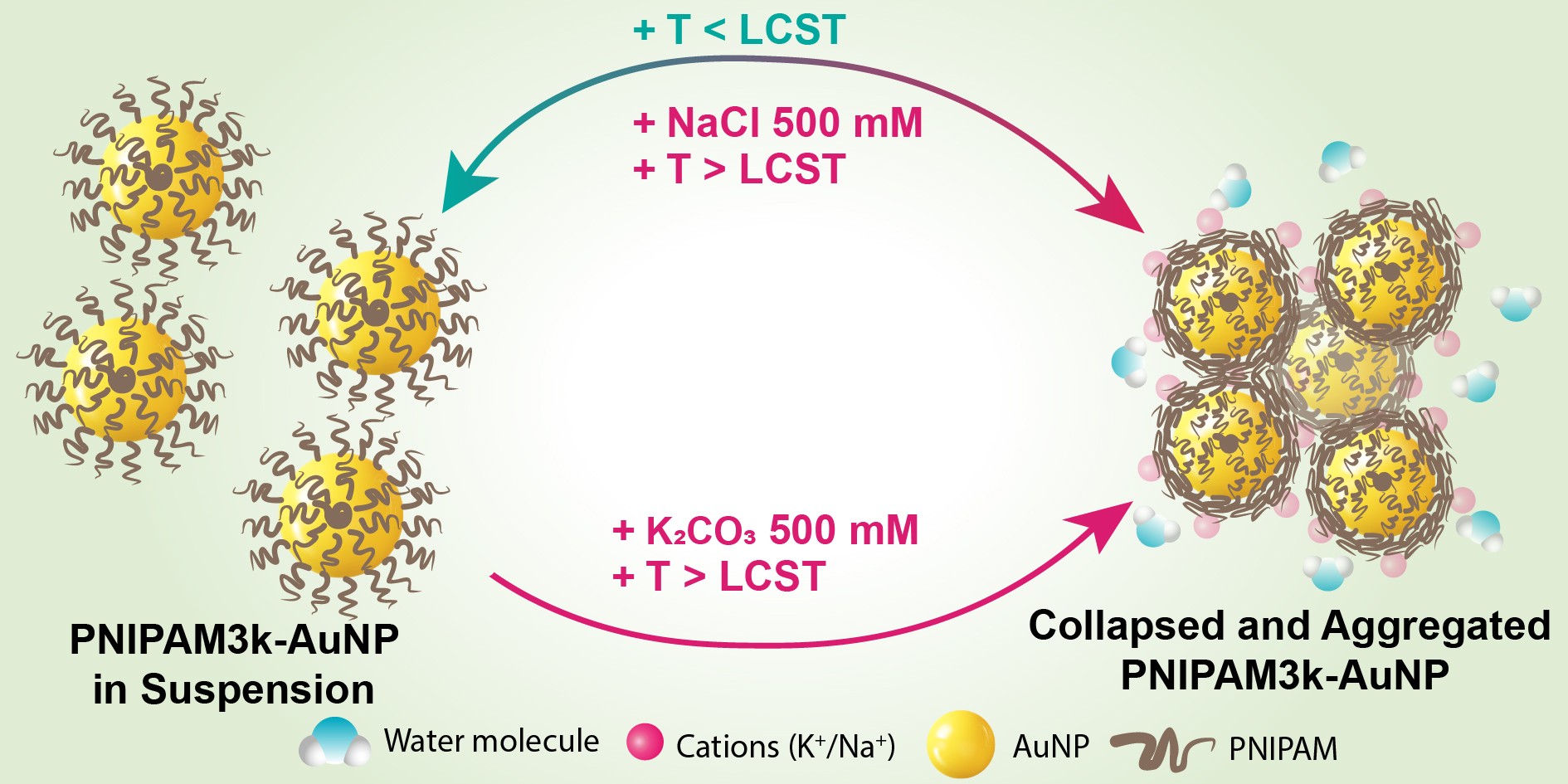}
    \label{toc}
\end{figure}

\clearpage
\onecolumn

\setcounter{page}{1}
\setcounter{figure}{0}
\setcounter{equation}{0}
\setcounter{table}{0}

\renewcommand{\thefigure}{S\arabic{figure}}
\renewcommand{\theequation}{S\arabic{equation}}
\renewcommand{\thetable}{S\arabic{table}}
\renewcommand{\thepage}{S\arabic{page}} 

\section{{\Large Supporting information}}
\begin{center} 
	{\textbf{\LARGE{Assembling PNIPAM-capped Gold Nanoparticles in Aqueous Solutions}}}\\
	\bigskip
	\normalsize
     Binay P. Nayak,$^\dagger$ 
    Hyeong Jin Kim,$^\dagger$
    Srikanth Nayak,$^{\dagger, \mid \mid}$
    Wenjie Wang, $^\ddagger$
    Wei Bu,$^\mathparagraph$ 
    Surya K. Mallapragada,$^*$$^,$$^\dagger$ and David Vaknin$^*$$^,$$^\S$\\
	\bigskip
	{$\dagger$\it Ames National Laboratory, and Department of Chemical and Biological Engineering, Iowa State University, Ames, Iowa 50011, United States}\\
	{$\ddagger$\it Division of Materials Sciences and Engineering, Ames National Laboratory, U.S. DOE, Ames, Iowa 50011, United States}\\
	{$\mathparagraph$\it NSF’s ChemMatCARS, Pritzker School of Molecular Engineering, University of Chicago, Chicago, Illinois 60637, United States}\\
	{$\S$\it Ames National Laboratory, and Department of Physics and Astronomy, Iowa State University, Ames, Iowa 50011, United States}\\
    {$\mid \mid$\it Current address: Chemical Sciences and Engineering Division, Argonne National Laboratory, Lemont, Illinois 60439, United States}\\
 
	\bigskip
	{E-mail: suryakm@iastate.edu; vaknin@ameslab.gov}\\
\end{center}

\section{Experimental Section}
\subsection{Preparation of Materials}
\begin{figure}[!ht]
 	\centering 
 	\includegraphics[width=0.5 \linewidth]{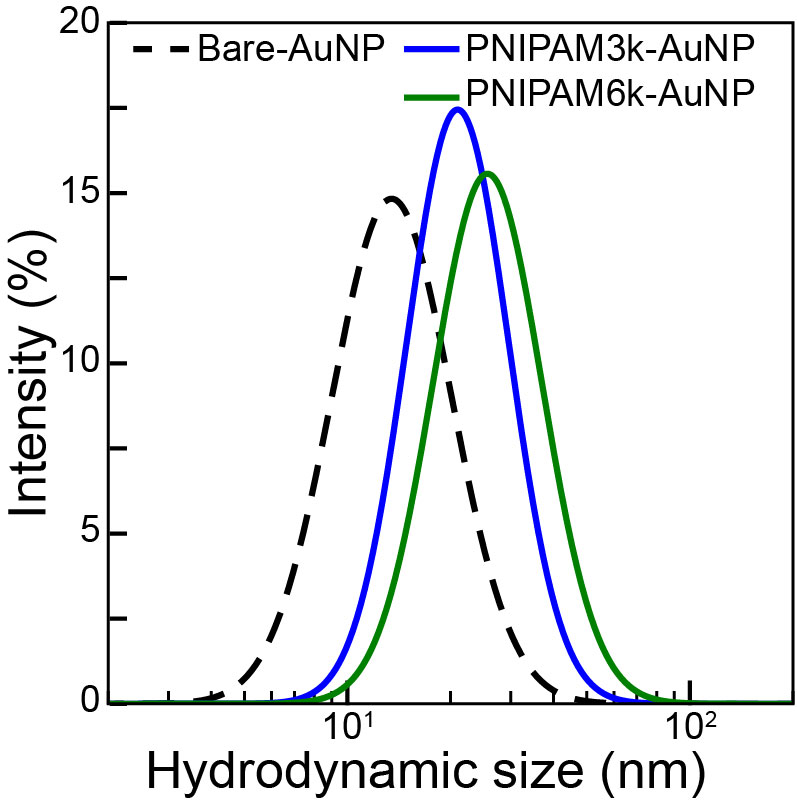}
 	\caption{DLS intensity percentage versus hydrodynamic size distribution for aqueous suspensions of bare surface AuNPs (dashed-line), PNIPAM3k-AuNPs (blue-solid line), and PNIPAM6k-AuNPs (green solid-line) at room temperature. The increase in hydrodynamic size confirms PNIPAM grafting.
 	}
 	\label{fig:dls}
 \end{figure}

Citrate stabilized AuNPs suspensions of nominal core diameter $\sim 10$ nm have been purchased from Ted Pella Inc. Thiolated PNIPAM of molecular weights (MW) $\sim 3$ and $\sim 6$ kDa have been purchased from Sigma Aldrich. The AuNPs were functionalized with thiolated PNIPAM (SH-PNIPAM) by a ligand exchange protocol, as outlined below.\cite{zhang2017macroscopic,minier2021poly} Thiolated PNIPAM ligands are dissolved in 50\% (v/v) ethanol solution and mixed with a calculated amount of AuNPs suspension (1:6000 AuNP to polymer ratio) continuously with a Roto-Shake Genie shaker (Scientific Industries, NY, USA) for at least 24 hours. The resulting PNIPAM-grafted AuNPs (PNIAPM-AuNPs) are then purified by centrifugation thrice at 20000 g for 1 hour to remove unbound ligands. Throughout the study, the term PNIAPM-AuNP is used generally to refer to PNIPAM-grafted AuNPs in general, while PNIAPM\textit{x}-AuNP indicates the PNIAPM ligands with an MW of \textit{x} grafted to AuNPs. For example, PNIAPM3k-AuNPs refer to AuNPs functionalized with 3 kDa MW of PNIPAM. The final concentration of the PNIPAM-grafted AuNPs is determined by measuring their absorbance using UV-vis spectroscopy (SpectraMax M3, Molecular Devices). For this study, the bare AuNPs are utilized at a concentration of approximately 9.46 nM, while the concentration of PNIPAM-AuNPs is adjusted to approximately 20 nM. DLS determines the percentage intensity distribution of the $D_{\rm H}$ of well-dispersed NPs in aqueous suspensions.  The measurements and data processing have been conducted with a NanoZS90  and its associated software, Zetasizer (Malvern, U.K.). From Figure \ref{fig:dls}, it is evident that $D_{\rm H}$ increases upon grafting and scales with the length of the polymer. We find that $D_{\rm H}$ $\simeq $ 12.1(3), 25.1(3), 27.8(6) nm for bare surface AuNPs, PNIPAM3k-AuNPs, and PNIPAM6k-AuNPs, respectively. This increase in size confirms the successful grafting of PNIPAM. For the X-ray measurements, the grafted AuNPs were studied at various electrolyte solutions such as  \ch{K2CO3} and  NaCl (Fisher Scientific) and PDAC (MW = 100 - 200 kDa, Sigma Aldrich). Prior to the X-ray experimentation, high concentrations of electrolyte stock solutions were prepared, and a calculated amount from the stock solution (relatively smaller than the volume of AuNPs suspension) was then mixed with the PNIPAM-AuNPs suspension and allowed to incubate for $\sim$ 20 minutes.

\subsection{X-ray Experimental Setup}
Synchrotorn-based SAXS measurements were conducted at beamline 12-ID-B, Advanced Photon Source (APS), Argonne National Laboratory. The experiment is performed in transmission mode with an incident X-ray energy of 13.3 keV. After incubation, The PNIPAM-AuNPs suspensions with the desired solvent condition are transferred to a thin-walled quartz capillary (inner diameter of 2 mm). The temperature of the capillary loaded with the sample was elevated from 20 to 80 \textdegree{C} and subsequently cooled down to 20 \textdegree{C} at a 2 \textdegree{C}/min rate. SAXS measurements were collected at each 10 \textdegree{C} increment interval. Details of the experimental setup, measurements, and data analysis can be found elsewhere. \cite{kim2021nanoparticleNTE,Nayak2019IPC} The normalized intensity ($S(Q)$) is obtained by dividing the reduced intensity of the SAXS profile by the corresponding measured form factor of individual NPs \cite{kim2021nanoparticleNTE}.

\clearpage 

\subsubsection{Additional DLS data}
\normalsize

\begin{figure}[!ht]
 	\centering 
 	\includegraphics[width= 0.8 \linewidth] 
     {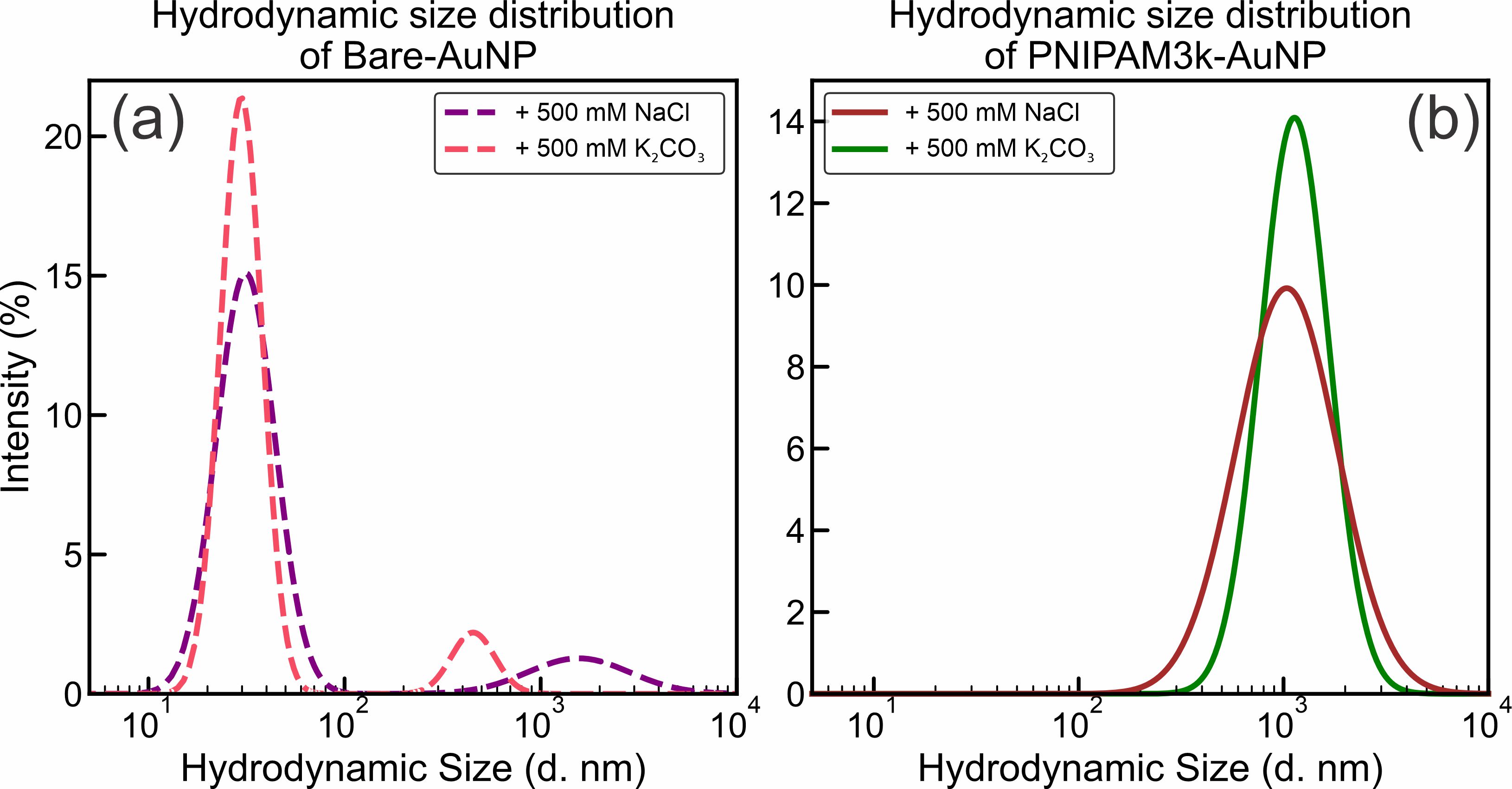}
 	\caption{Hydrodynamic size distribution of (a) Bare AuNPs and (b) PNIPAM3k-AuNPs in the presence of 500 mM NaCl and \ch{K2CO3} at room-temperature as indicated. The bare AuNPs show minimal aggregation, whereas PNIPAM3k-AuNPs spontaneously aggregate at room temperature in the presence of salts.
 	}
 	\label{DLS_3k_salts}
 \end{figure} 

\begin{figure}[!ht]
 	\centering 
 	\includegraphics[width= 0.8 \linewidth]{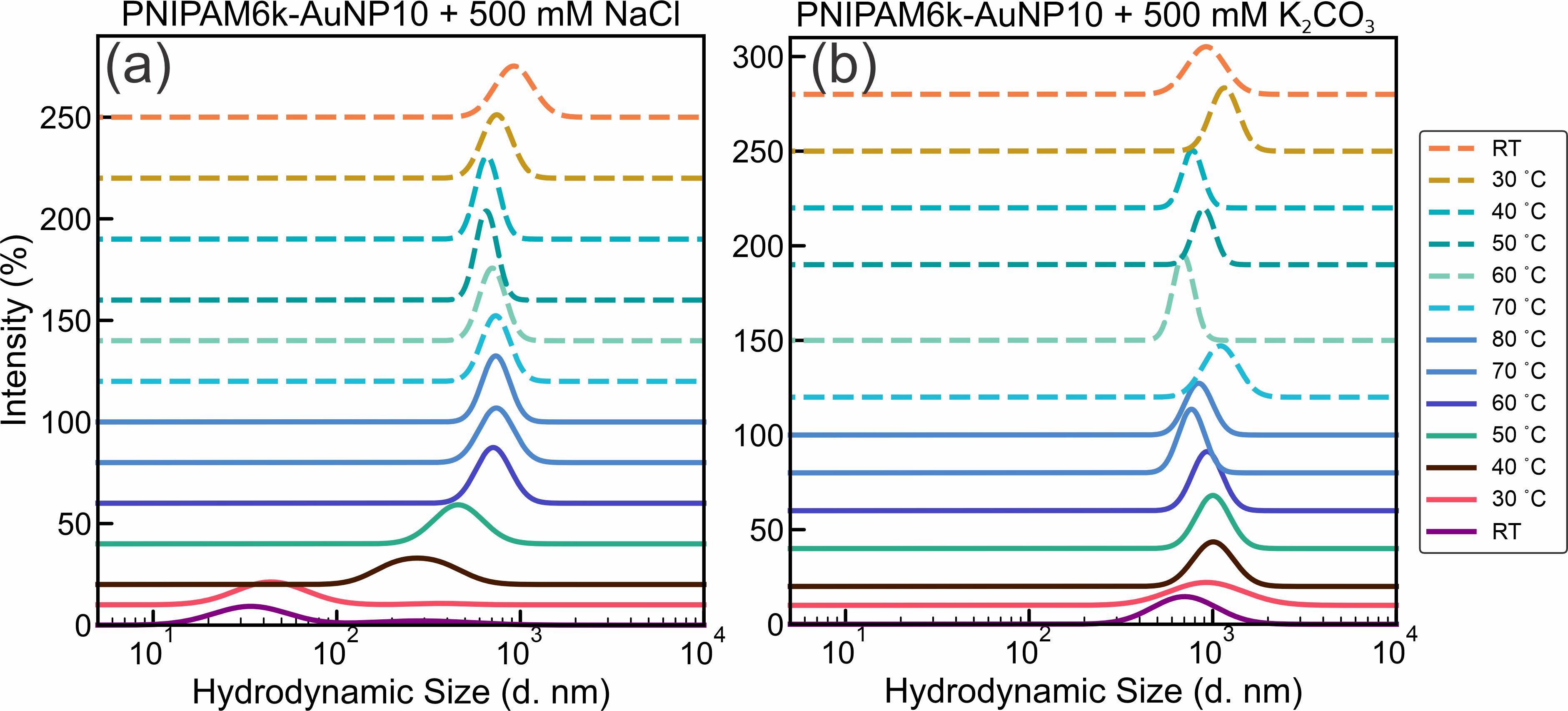}
 	\caption{Hydrodynamic size distribution of PNIPAM6k-AuNPs in the presence of (a) 500 mM NaCl and (b) 500 mM \ch{K2CO3} at various temperatures as indicated. 
 	}
 	\label{DLS_6k_salts}
 \end{figure}

\noindent Figure \ref{DLS_3k_salts} shows DLS measurements for (a) bare AuNP and (b) PNIPAM3k-AuNP in the presence of 500 mM NaCl and \ch{K2CO3} at room-temperature as indicated. In the case of bare AuNPs, a tendency towards partial aggregation is observed under both salt conditions, although a significant proportion remains non-aggregated. In contrast, the PNIPAM3k-AuNPs exhibit a more pronounced response in the presence of salt, showing spontaneous aggregation when exposed to both NaCl and \ch{K2CO3}. The aggregates are considered as amorphous suggested by SAXS results. Figure \ref{DLS_6k_salts} displays DLS measurements for PNIPAM6k-AuNP at (a) 500 mM NaCl and (b) 500 mM \ch{K2CO3} at various temperatures as indicated. At room temperature, PNIPAM6k-AuNP in 500 mM NaCl does not aggregate. However, increasing temperature above the LCST (to 40 \textdegree{C}) induces aggregation, which intensifies progressively with further temperature increase to 80 \textdegree{C}. This phenomenon is rationalized by the fact that PNIPAM corona collapses on the core, and binding is facilitated by the salts and thus promotes aggregation.  When the temperature is reduced from 80 \textdegree{C} back to room temperature, the PNIPAM6k-AuNPs remain aggregated. Interestingly, the aggregation size at room temperature after heating is larger than at 80 \textdegree{C}. This suggests that while the polymer corona may partially revert to its pre-heating expanded state, the nanoparticles remain as amorphous aggregates. In the presence of 500 mM \ch{K2CO3}, PNIPAM6k-AuNPs show spontaneous amorphous aggregation at room temperature. Intriguingly, upon gradually heating the suspension above the LCST, there is an observable decrease in aggregate size. This reduction is likely due to the shrinking of the PNIPAM corona (collapse) at elevated temperatures, reducing inter-particle distances. Cooling the suspension reverses this effect, leading to an increase in aggregate size.

 \begin{figure}[!ht]
 	\centering 
 	\includegraphics[width= 0.5 \linewidth]{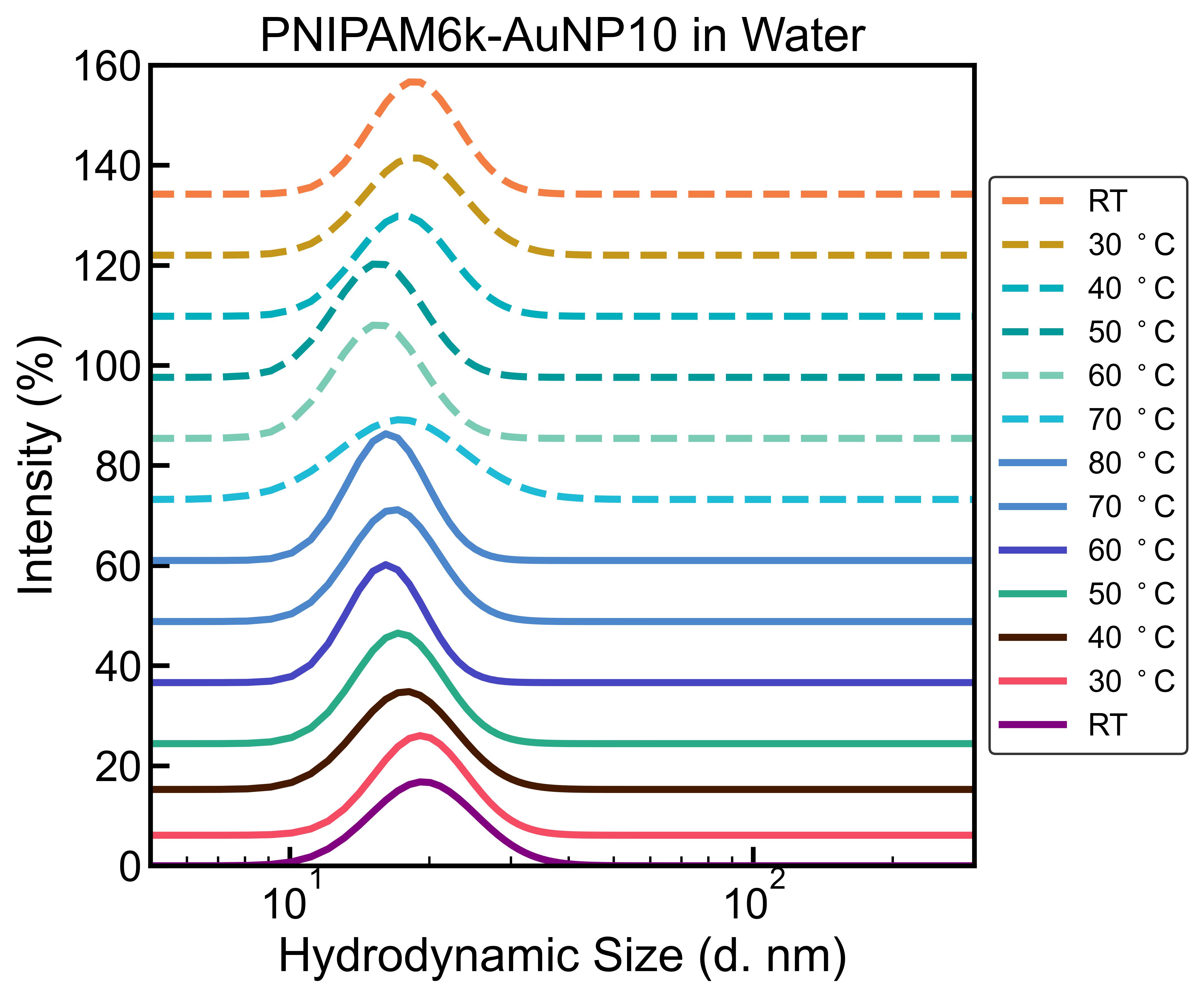}
 	\caption{Hydrodynamic size distribution of PNIPAM6k-AuNPs in water at various temperatures as indicated. Without any electrolytes, PNIPAM6k-AuNPs do not aggregate upon heating the suspension.
 	}
 	\label{DLS_6k_water}
 \end{figure} 

As a control experiment, we examine the effect of temperature in the absence of salt on PNIPAM6k-AuNP; we conduct DLS measurements as shown in Figure \ref{DLS_6k_water}. The message from this figure is that temperature by itself does not induce aggregation even above the LCST, and salt is needed to achieve aggregation, as discussed above. It is interesting to note that, as expected, raising the temperature changes the $D_{\rm H}$ to smaller values in accordance with the known behavior of the grafted polymer above the LCST \cite{wu1998globule}. We note that the conditions under which we conduct in-situ SAXS measurements and DLS measurements are not the same, and some variations are expected.
 
\subsubsection{Additional SAXS data}
\normalsize

\begin{figure}[!ht]
  \centering
  \begin{minipage}{.45\textwidth}
        \centering 
 	\includegraphics[width= \linewidth]{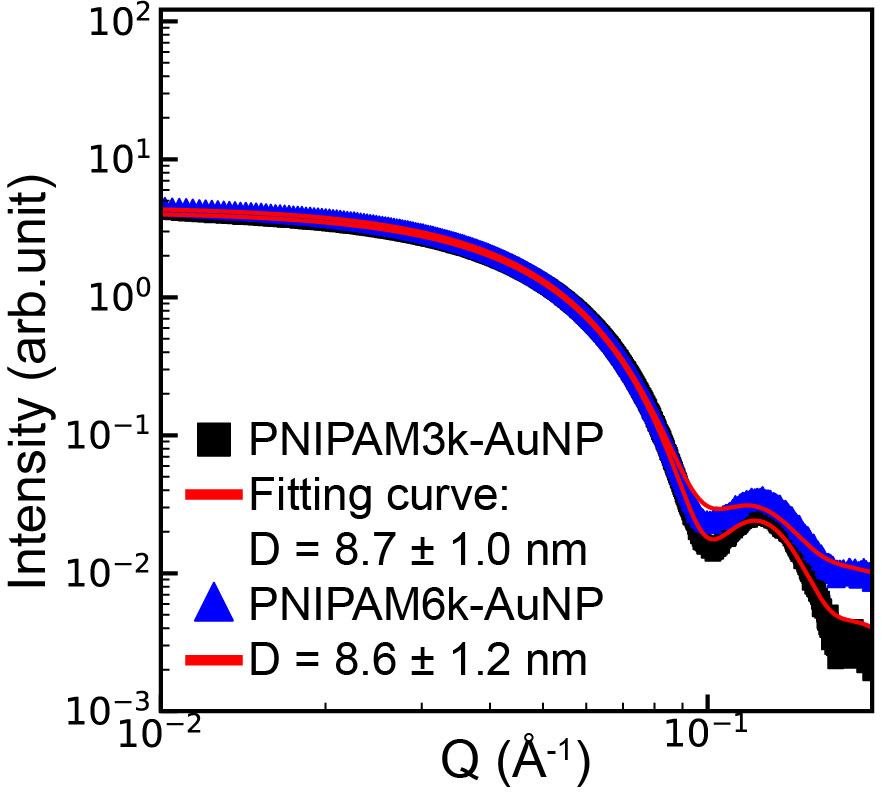}
 	\caption{SAXS data for PNIPAM3k-AuNPs and PNIPAM6k-AuNPs in water (i.e., without any salts at room temperature) and their best-fit curves for the form factor profile of sphere (red solid lines).
 	}
 	\label{form_fit}
  \end{minipage}%
    \hspace{0.05\textwidth} 
    \begin{minipage}{.45\textwidth}
 	\centering 
 	\includegraphics[width= \linewidth]{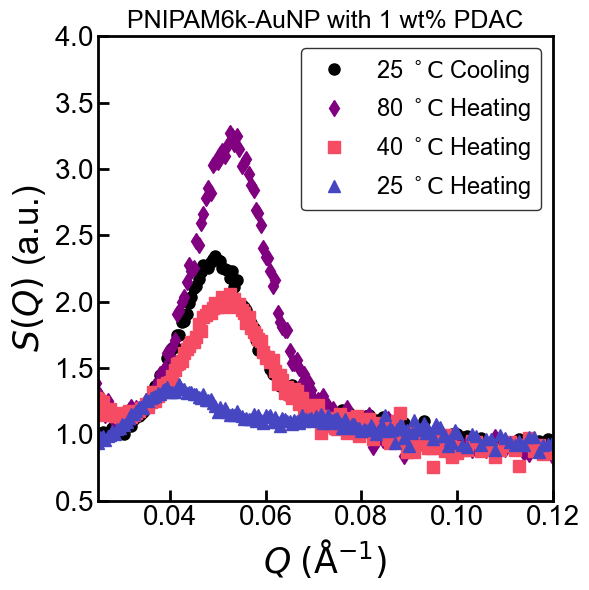}
 	\caption{Normalized intensity profile $S(Q)$ profile obtained at different temperatures for PNIPAM6k-AuNPs at $\sim$ 1wt $\%$ of Poly(diallyl-dimethylammonium chloride)}
 	\label{PNIPAM_PDAC}
  \end{minipage}
 \end{figure}

Figure \ref{form_fit} shows SAXS data from PNIPAM3k-AuNPs and PNIPAM6k-AuNPs in water (i.e., without any salts at room temperature) and their best-fit curves for the form factor profile of sphere (red solid lines). We note that due to the lack of electron density (ED) contrast between the polymer and the aqueous medium, the X-ray scattering from the grafted nanoparticle is dominated by the gold nanoparticle (AuNP) core.

Figure \ref{PNIPAM_PDAC} shows normalized intensity profile $S(Q)$ profile obtained at different temperatures for PNIPAM6k-AuNPs at $\sim$ 1wt $\%$ of Poly(diallyl-dimethylammonium chloride) (PDAC). The polymer, unlike the salts, induces assembly at room temperature with a broad peak at $Q_{0} = 0.04$ \AA $^{-1}$ which, upon raising the temperature above 35 \textdegree{}C, shifts to $Q_{0} = 0.053$ \AA $^{-1}$. The process is irreversible; namely, the peak at room temperature remains pronounced. 

\begin{figure}[!ht]
 	\centering 
 	\includegraphics[width= 0.8 \linewidth]{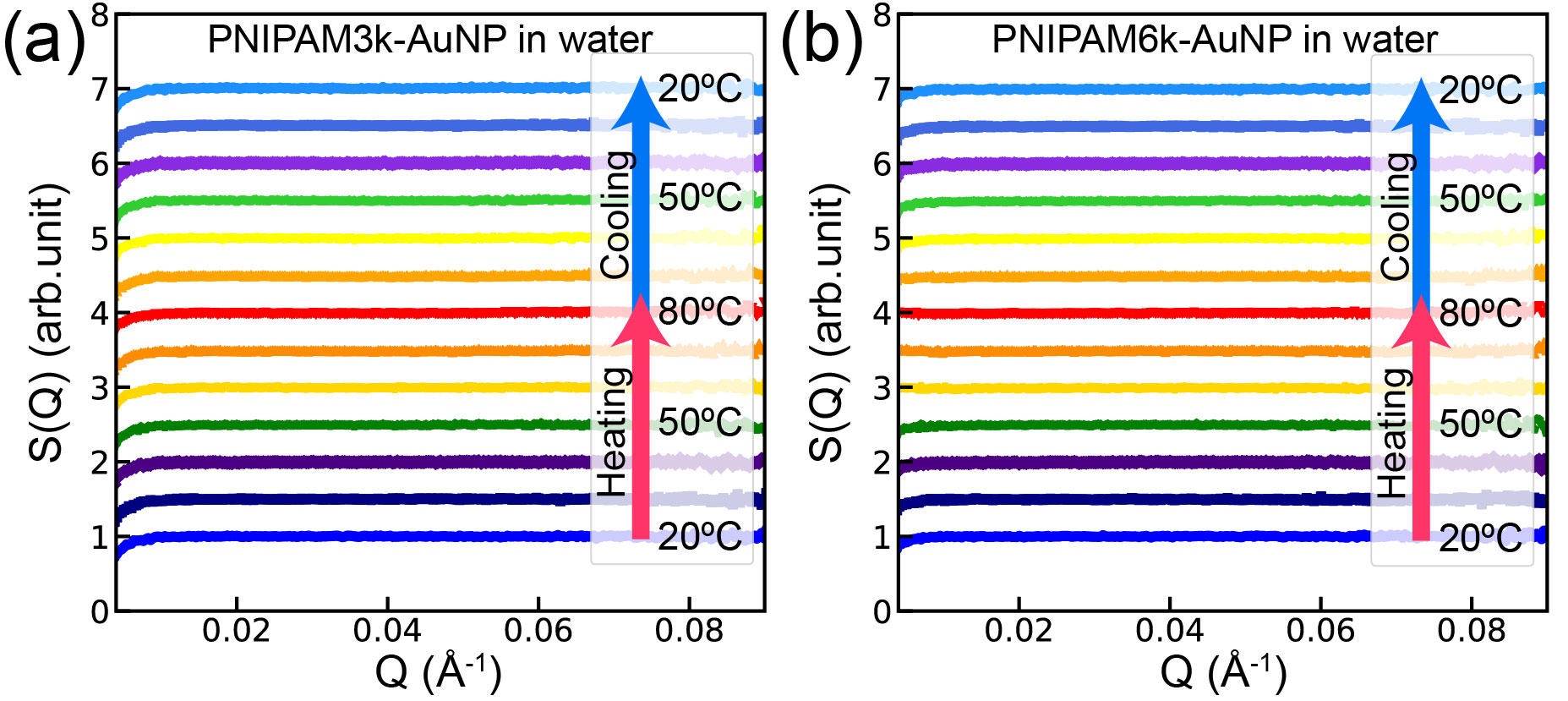}
 	\caption{Normalized intensity profiles ($S(Q)$) from SAXS data for (a) PNIPAM3k-AuNP and (b) PNIPAM6k-AuNP in water under the temperature controls.
 	}
 	\label{form_temp_s}
 \end{figure} 

 \begin{figure}[!ht]
 	\centering 
 	\includegraphics[width= 0.8 \linewidth]{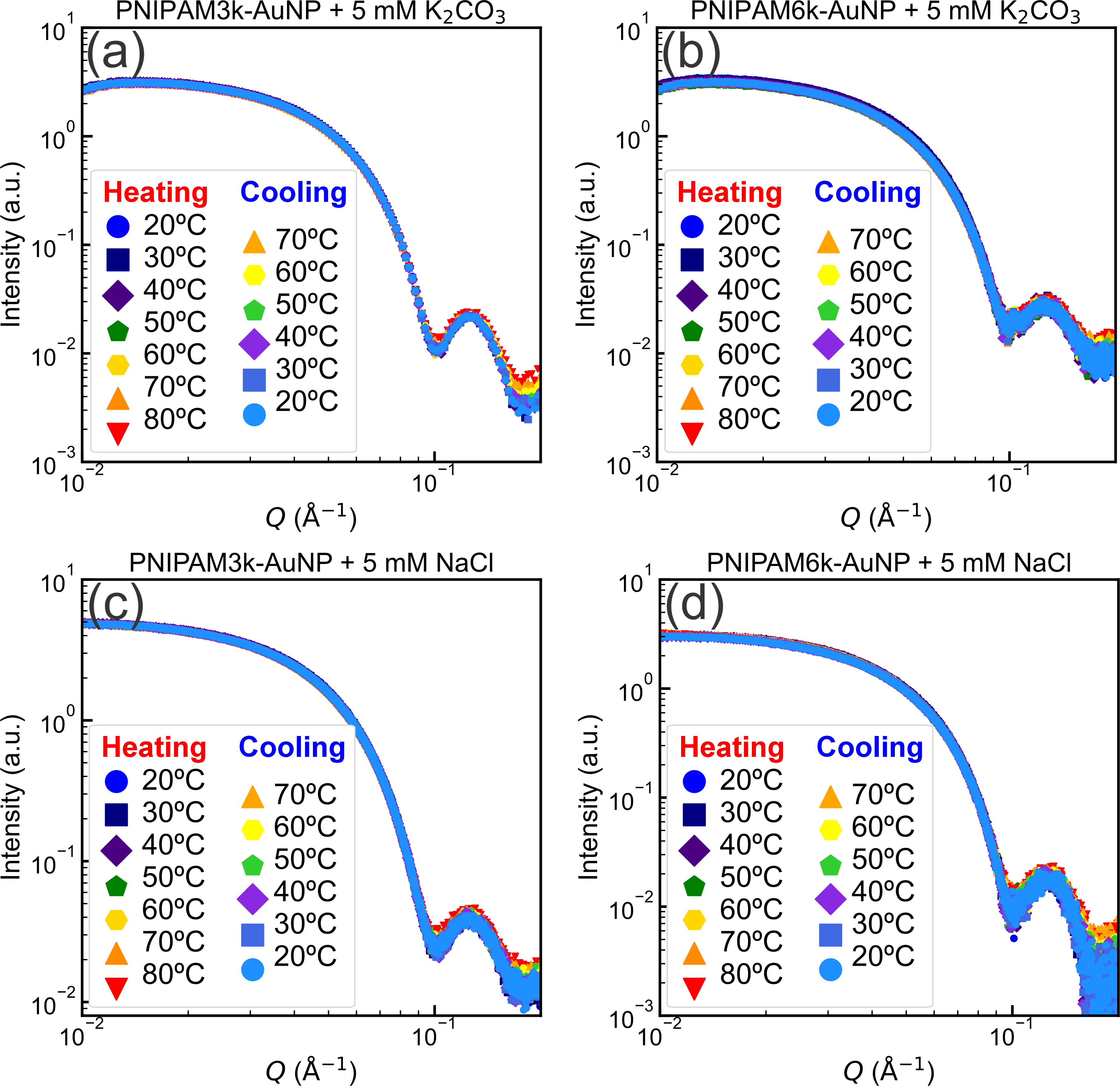}
 	\caption{SAXS data for PNIPAM3k-AuNPs with (a) 5 mM of \ch{K2CO3} and (c) 5 mM NaCl and SAXS data for PNIPAM6k-AuNP with (b) 5 mM of \ch{K2CO3} and (d) 5 mM NaCl under the control of temperatures in the range of  $20-80$ \textdegree{}C heating and $80-20$ \textdegree{}C cooling.
 	}
 	\label{3D_lowsalt}
 \end{figure} 

Figure \ref{form_temp_s} shows normalized intensity profiles ($S(Q)$) from SAXS data for (a) PNIPAM3k-AuNP and (b) PNIPAM6k-AuNP in water under the temperature controls. The constant values of $S(Q)$ indicate dispersed NPs in the suspensions even above the LCST. Adding small amounts of salts 5 mM NaCl or \ch{K2CO3} (Figure \ref{3D_lowsalt}) yields results that are practically the same as those in water. Note that for deductive purposes, we show the un-normalized scattering, namely the form factor of the NPs. 

 \begin{figure}[!ht]
 	\centering 
 	\includegraphics[width= \linewidth]{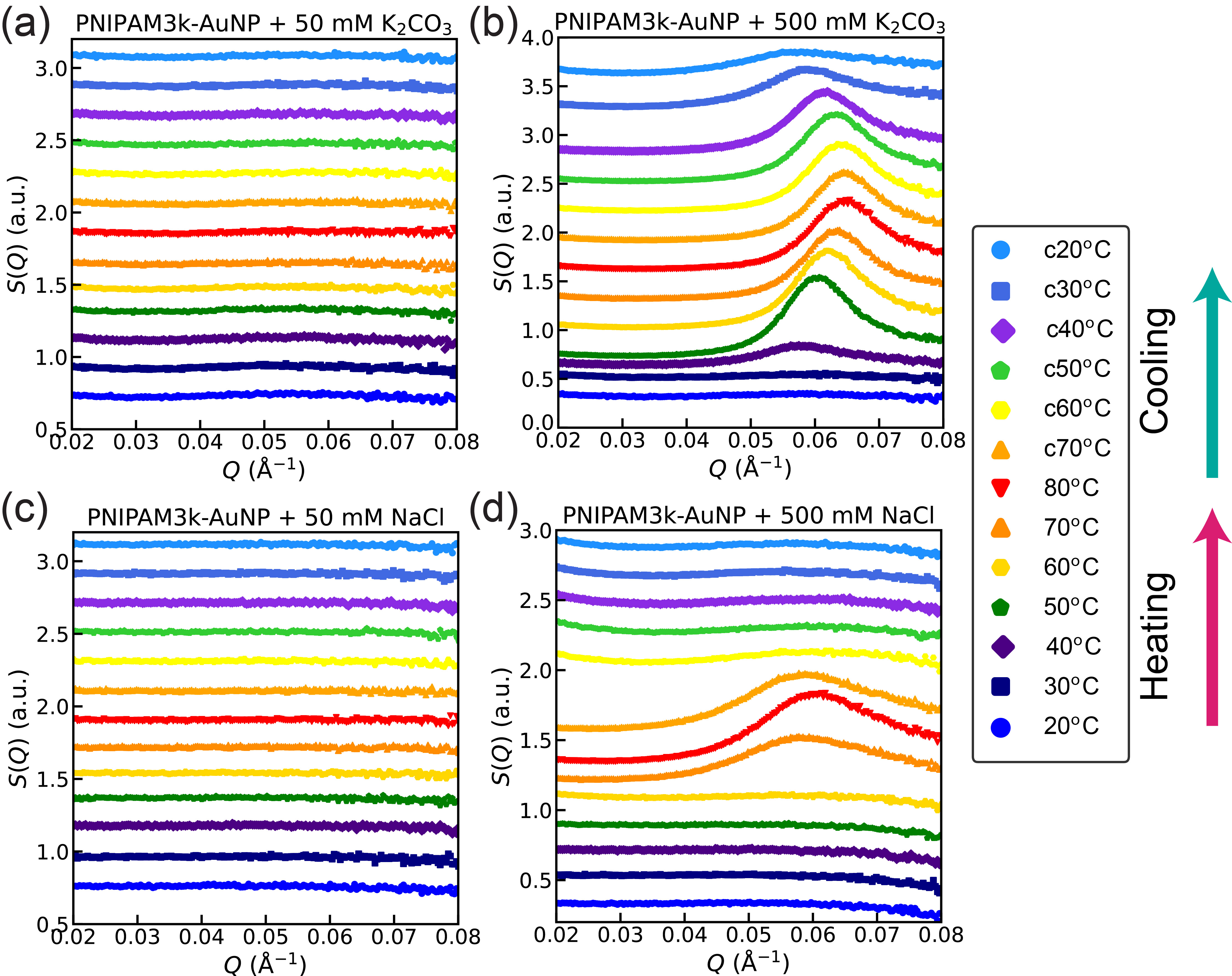}
 	\caption{Normalized intensity profile $S(Q)$ profile obtained at different temperatures for PNIPAM3k-AuNPs at (a) 50 mM \ch{K2CO3} (b) 500 mM \ch{K2CO3} (c) 50 mM NaCl (d) 500 mM NaCl}
 	\label{3k_si}
 \end{figure} 

\clearpage 

 \begin{table*}[ht] 
\centering
\caption{\small Summary of the nearest-neighbor distance and FWHM for different PNIPAM-AuNP assemblies induced under various electrolytes and temperatures as indicated. Temperature values labeled with the prefix “c” indicate that they were performed after cooling from 80 \textdegree{}C.}
\label{tbl:lattice}
\begin{threeparttable}
\small
\begin{tabular*}{\textwidth}{@{\extracolsep{\fill}}cccccc}
    \toprule
    Name of & Suspensions condition & Temperature & Q$_{\rm center}$& $d_{\rm NN}^{**}$ & FWHM$^\#$ \\
    PNIPAM-AuNP & ($^\circ$C) & & (\AA$^{-1}$)& (nm) & (\AA$^{-1}$)\\
    \midrule
    \multirow{14}{*}{PNIPAM3k-AuNP}
    & + \ch{K2CO3} 50 mM & \multicolumn{4}{c}{\textit{No structure observed at any temperature}} \\
    \cmidrule{2-6}
    
    & + NaCl 50 mM & \multicolumn{4}{c}{\textit{No structure observed at any temperature}} \\
    \cmidrule{2-6}
    
    & \multirow{9}{*}{+ \ch{K2CO3} 500 mM} & 40 & -- & -- & --\\
    & & 50 & 0.0606 (1) & 10.4 (1) & 0.0062 (2) \\
    & & 60 & 0.0624 (1)& 10.1 (1) & 0.0069 (1) \\
    & & 70 & 0.0636 (1)& 9.9 (1) & 0.0067 (1) \\
    & & 80 & 0.0649 (1)& 9.7 (1) & 0.0066 (1) \\
    & & c70 & 0.0648 (1)& 9.7 (1) & 0.0066 (1) \\
    & & c60 & 0.0643 (1)& 9.8 (1) & 0.0067 (1) \\
    & & c50 & 0.0635 (1)& 9.9 (1) & 0.0069 (1) \\
    & & c40 & 0.0619 (1)& 10.2(1) & 0.0075 (2) \\
    
    \cmidrule{2-6}
    
    & \multirow{3}{*}{+ NaCl 500 mM} & 70 & 0.0596 (1)& 10.6 (2) & 0.0161 (2)\\
    & & 80 & 0.0611 (1)& 10.3 (2) & 0.0112 (2)\\
    & & c70 & 0.0596 (1)& 10.6 (2) & 0.0141 (2)\\
    
    \midrule
    
   \multirow{20}{*}{PNIPAM6k-AuNP} & \multirow{5}{*}{+ \ch{K2CO3} 50 mM$^*$} & 60 & 0.0314 (1)& 15.0 (7) & 0.0213 (7) \\
    & & 70 & 0.0324 (1)& 14.5 (4) & 0.0212 (2) \\
    & & 80 & 0.0359 (1)& 13.1 (4) & 0.0291 (1) \\
    & & c70 & 0.0364 (1)& 12.9 (4) & 0.0213 (2) \\
    & & c60 & 0.0317 (1)& 14.9 (2) & 0.0174 (2) \\
    
    \cmidrule{2-6}
    
    & \multirow{5}{*}{+ NaCl 50 mM} & 60 & 0.0602 (1)& 10.4 (1) & 0.0218(4) \\
    & & 70 & 0.0587 (1)& 10.7 (1) & 0.0095 (1) \\
    & & 80 & 0.0577 (1)& 10.9 (1) & 0.0069 (1) \\
    & & c70 & 0.0578 (1)& 10.9 (1) & 0.0076 (1) \\
    & & c60 & 0.0597 (1)& 10.5 (1) & 0.0122 (2) \\

    \cmidrule{2-6}

    & + \ch{K2CO3} 500 mM & \multicolumn{4}{c}{\textit{No structure observed at any temperature}} \\

    \cmidrule{2-6}
    
    & \multirow{9}{*}{+ NaCl 500 mM} & 40 & -- & -- & --\\
    & & 50 & 0.0541 (1)& 11.6 (1) & 0.0089 (1) \\
    & & 60 & 0.0548 (1)& 11.5 (1) & 0.0057 (1) \\
    & & 70 & 0.0550 (1)& 11.4 (1) & 0.0048 (1)\\
    & & 80 & 0.0549 (1)& 11.4 (1) & 0.0045 (1) \\
    & & c70 & 0.0550 (1)& 11.4 (1) & 0.0045 (1) \\
    & & c60 & 0.0548 (1)& 11.5 (1) & 0.0047 (1) \\
    & & c50 & 0.0537 (1)& 11.7 (1) & 0.0056 (1) \\
    & & c40 & 0.0541 (1)& 11.6 (1) & 0.0142 (6) \\
    
    \bottomrule
\end{tabular*}
\begin{tablenotes}
    \item[*] {\small Only first peak corresponding to \{111\} plane of diamond cubic structure is reported. neighbor distance ($d_{\rm NN}$) for diamond cubic lattice is  $\frac{3\pi}{2Q_{111}}$.}
    \item[**] {\small $Q_{\rm center}$, $d_{\rm NN}$ and FWHM is reported for structures where applicable.}
    \item[\#] {\small FWHM - Full Width at Half Maximum is estimated from the Lorentzian fitting function of the prominent peak.}
\end{tablenotes}
\end{threeparttable}
\end{table*}

\clearpage 

\subsubsection{Rationalization of Diamond-like structure for PNIPAM6k-AuNPs in 50 mM \ch{K2CO3} at 80 \textdegree{}C}

\begin{figure}[!h]
 	\centering 
 	\includegraphics[width=   \linewidth]{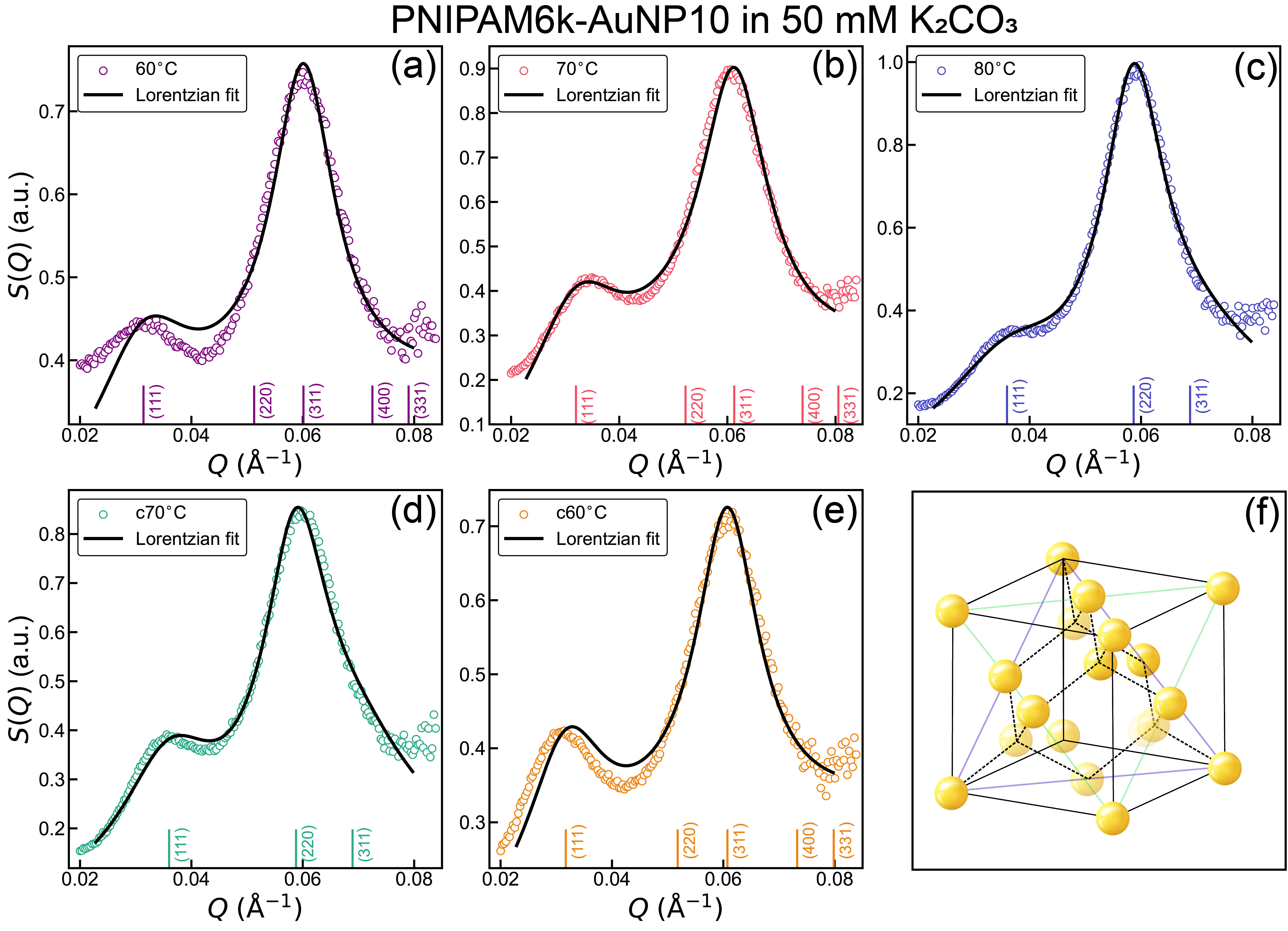}
 	\caption{ (a-e) Normalized $S(Q)$ profiles of PNIPAM6k-AuNPs in 50 mM \ch{K2CO3} across various temperatures as indicated. The solid lines represent fits using Lorentzian-shaped functions based on a diamond-cubic lattice model. Vertical lines mark peak positions with their respective Miller indices. Temperatures with a preceding 'c' denote cooling conditions. For clarity, plots have been vertically shifted. (f) Schematic illustration of a simple diamond-cubic lattice. }
 	\label{diamond}
\end{figure}

Below, we examine a few model structures for SAXS measurements of PNIPAM6k-AuNPs in 50 mM \ch{K2CO3} at 80 $^{\circ}$C and show that the more-likely structure is diamond-like, albeit at very short-range order (correlation-length is on the order of 1-2 unit cells). Figure \ref{fig:lattice_compare} shows the SAXS data with expected intensities at the various Bragg reflections indicated and scaled as vertical dashed lines. See Table \ref{table:multiplicity} for more detailed comparisons of relative calculated intensities. It is clear that for the BCC, FCC, and HCP model structures, the ratio between the first and higher-order peaks is greater than one and smaller than one for the diamond-like structure, as observed experimentally. We have, in fact, modeled these data sets using the four models, and the best fit is obtained with the diamond-like structure. Figure \ref{diamond} shows fits the data at various temperatures, relaxing the calculated intensity $\pm$ 20\%. The parameters are listed in Table \ref{table:diamond_param}. We note that our model is based on the similarity of our diffraction pattern with that observed in Ref.[\!\! \citenum{kalsin2006electrostatic}] . Although the assignment to a diamond structure should be taken with reservations, it is consistent with the conformation of PNIPAM at high temperatures; namely, the polymer is collapsed.

\vspace{12pt} 

\noindent 
\begin{minipage}{0.5\textwidth} 
\centering
\begin{tabular}{|c| c c c c|}
\hline
&$(h,k,l)$ & $Q_{hkl}/Q_0$ & $m_{hkl}$ & $I_{cal}$ \\ 
\toprule
\hline
\multirow{7}{*}{\rotatebox[origin=c]{90}{$bcc$}}
& $(1,1,0)$ & 1 & 12 & 100.0\\
&$(2, 0, 0)$ & $\sqrt{2}$ & 6 & 25.0\\
&$(2, 1, 1)$ & $\sqrt{3}$ & 24& 66.7\\
&$(2, 2, 0)$ &2 & 12 & 25.0\\
&$(3, 1, 0)$ & $\sqrt{5}$ &24 & 40.0\\
&$(2, 2, 2)$ & $\sqrt{6}$ &8 & 11.1\\
&$(3, 2, 1)$ & $\sqrt{7}$ &48 & 57.1\\
\cline{1-5}
\multirow{5}{*}{\rotatebox[origin=c]{90}{$fcc$}}
&$(1, 1, 1)$&  1  &  8 & 100.0\\
&$(2, 0, 0)$ & $\sqrt{4/3}$ & 6 & 56.3\\
&$(2, 2, 0)$ & $\sqrt{8/3}$ & 12 & 56.3 \\
&$(3, 1, 1)$ & $\sqrt{11/3}$ &  24 & 81.8\\
&$(2, 2, 2)$ &  2  &  8 & 25.0\\
\cline{1-5}
\multirow{7}{*}{\rotatebox[origin=c]{90}{hexagonal}}
&$(1,0,0)$ & 1 & 6 & 100.0\\
&$(1,1,0)$ & $\sqrt{3}$ & 6 & 33.3\\
&$(2,0,0)$ & 2 & 6 & 25.0\\
&$(2,1,0)$ & $\sqrt{7}$ &12 &28.6\\
&$(3,0,0)$ & 3 & 6 & 11.1\\
&$(2,2,0)$ & $\sqrt{12}$ & 6 &8.3\\
&$(3,1,0)$ & $\sqrt{13}$ & 12 & 15.4\\
\cline{1-5}
\multirow{4}{*}{\rotatebox[origin=c]{90}{diamond}}
&$(1, 1, 1)$ & 1& 8 & 88.9 \\
&$(2, 2, 0)$ & $\sqrt{8/3}$ & 12  &100.0 \\
&$(3, 1, 1)$ & $\sqrt{11/3}$ & 24 & 72.7\\
&$(4, 0, 0)$ & $\sqrt{16/3}$ & 6 & 25.0\\

\hline
\end{tabular}
\captionof{table}{Multiplicity ($m_{hkl}$) and relative observed intensity ($I_{cal}$) for different crystal structures.}
\label{table:multiplicity}
\end{minipage}
\hfill 
\begin{minipage}{0.45\textwidth} 
\centering
\includegraphics[width=\linewidth]{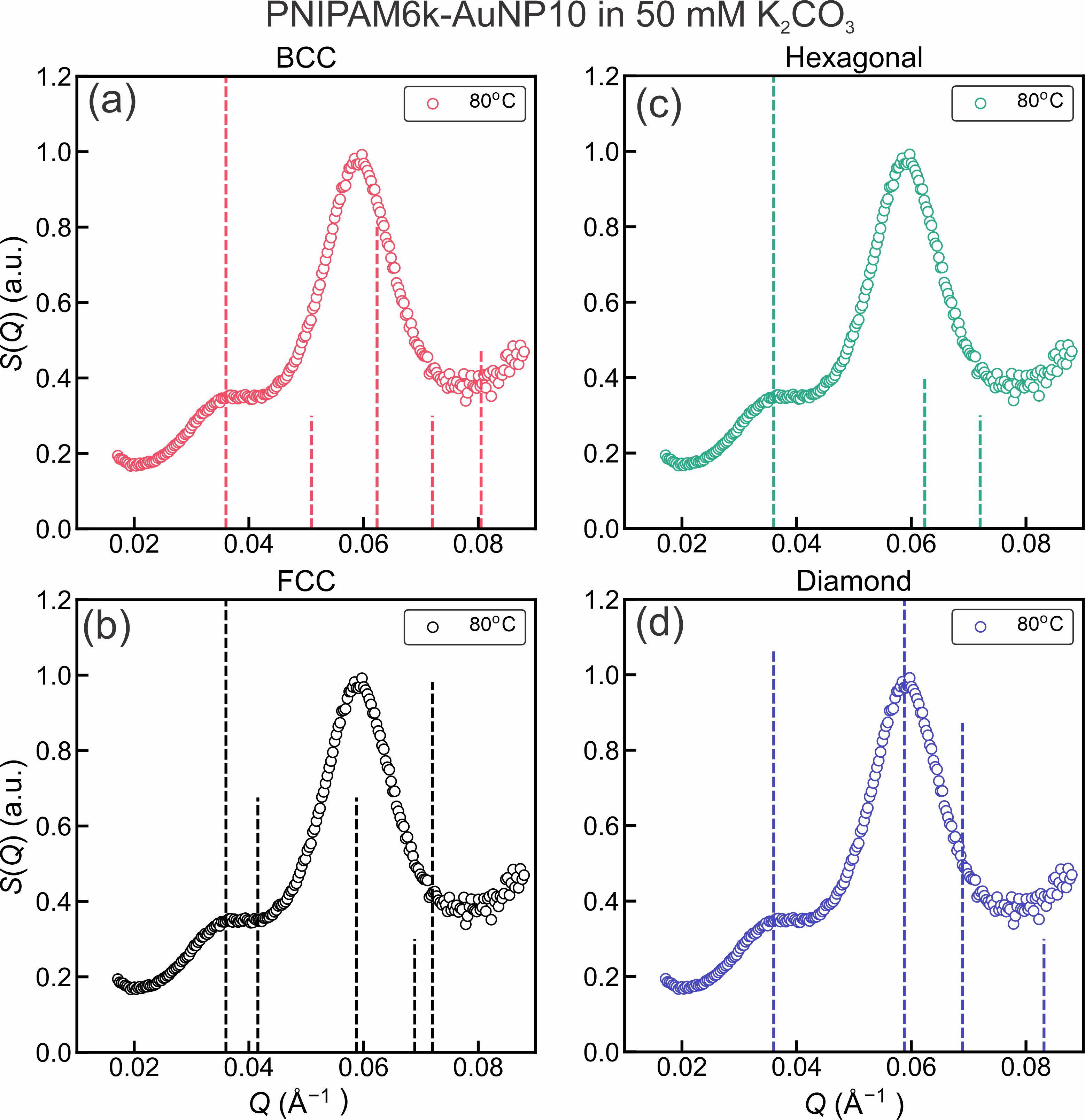}
\captionof{figure}{Calculated peak positions derived from $Q_{0} = 0.0359$ \AA$^{-1}$ (Table \ref{tbl:lattice}) for various lattice types are indicated by dashed lines. The height of each line corresponds to the relative calculated intensities for the respective lattice, as detailed in Table \ref{table:multiplicity}.}
\label{fig:lattice_compare}
\end{minipage} 

\vspace{12pt} 

The SAXS intensity $I_{hkl}$, i.e., integrated area for each indexed $(h,k,l)$ peak, is mainly proportional to the corresponding multiplicity, $m_{hkl}$, multiplied by Lorentz factor that is proportional to $1/Q^2$, accounting for the observed intensity for powder diffraction. Here, we list $m_{hkl}$ and relative calculated intensity, $I_{cal}$, that is proportional to $m_{hkl}/Q^2$. $Q_{0}$ is the primary peak position, and the maximum peak intensity is scaled to 100. Within our $Q$ range $0.01-0.08$ {\AA}$^{-1}$, only the diamond structure is the most likely to account for the two apparent intensity maxima by three designated indexed peaks with comparable intensities.

\begin{table}[H]
\centering
\small
\begin{tabular}{cccc|ccc|ccc}
\toprule
\multirow{3}{*}{($h,k,l$)} & \multicolumn{3}{c}{60 \textdegree{}C} & \multicolumn{3}{c}{70 \textdegree{}C} & \multicolumn{3}{c}{80 \textdegree{}C} \\
\cmidrule(lr){2-4} \cmidrule(lr){5-7} \cmidrule(l){8-10}
& $Q_{hkl}$ & Intensity & FWHM & $Q_{hkl}$ & Intensity & FWHM & $Q_{hkl}$ & Intensity & FWHM \\
& (\AA$^{-1}$) & (a.u.) & (\AA$^{-1}$) & (\AA$^{-1}$) & (a.u.) & (\AA$^{-1}$) & (\AA$^{-1}$) & (a.u.) & (\AA$^{-1}$) \\
\midrule
(1,1,1) & 0.0314 & 4.3 & 0.0213 & 0.0324 & 4.3 & 0.0212 & 0.0359 & 4.3 & 0.0291   \\
(2,2,0) & 0.0512 & 4.5 & 0.0505 & 0.0529 & 4.5 & 0.0445 & 0.0586 & 6.5 & 0.0132   \\
(3,1,1) & 0.0601 & 6.4 & 0.0134 & 0.0619 & 6.4 & 0.0149 & 0.0688 & 5.0 & 0.0350   \\
\bottomrule
\end{tabular}
\caption{Fitting parameters extracted from Lorentzian functions for the diamond-like cubic lattice of PNIPAM6k-AuNP10 at 50 mM \ce{K2CO3} under different temperature conditions.}
\label{table:diamond_param}
\end{table}
 
\clearpage
\subsubsection{2D XRR and GISAXS data}
 \begin{figure*}[!ht]
 	\centering 
 	\includegraphics[width=.9 \linewidth]{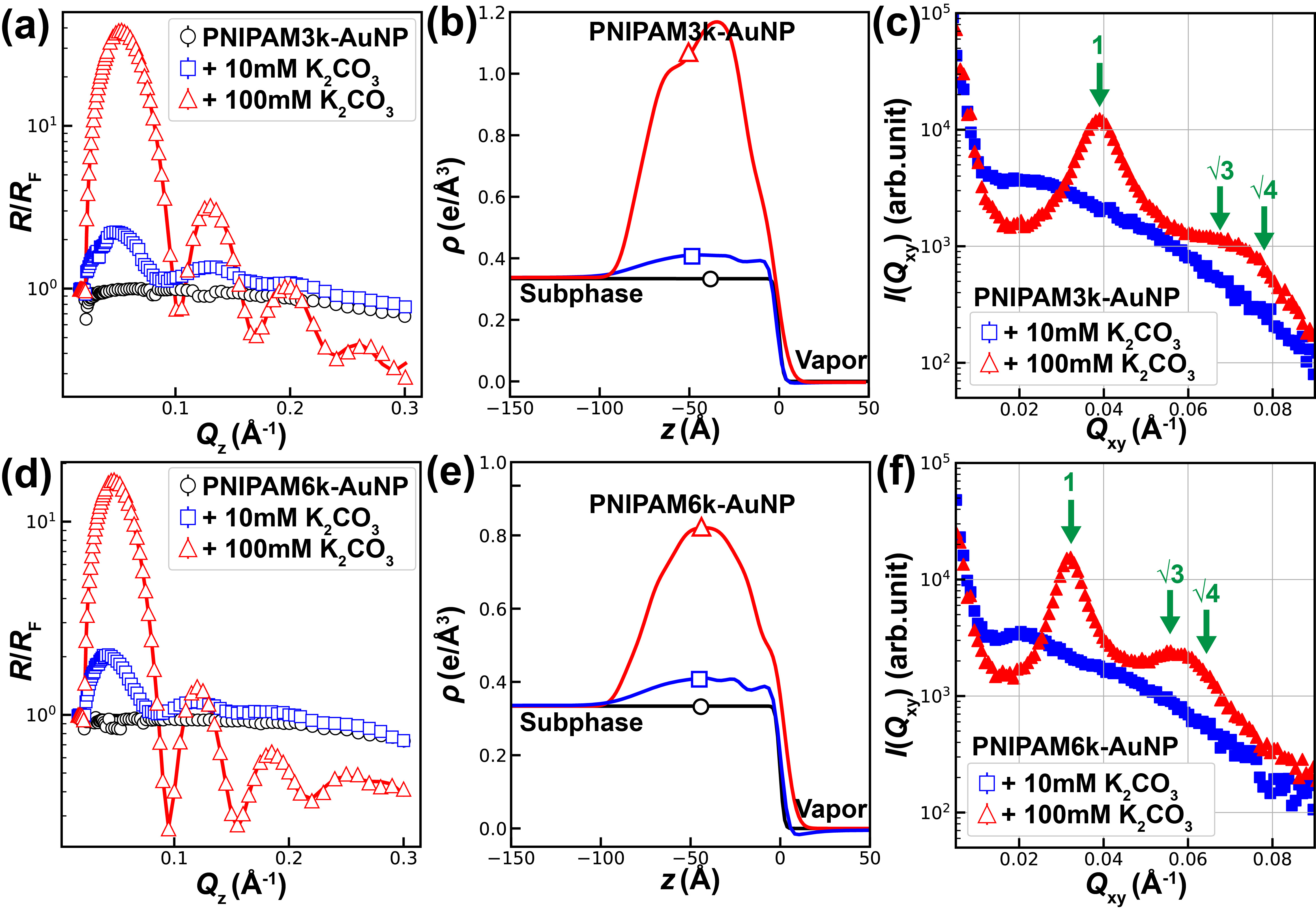}
 	\caption{ $R/R_F$ data for (a) PNIPAM3k-AuNP and (d) PNIPAM6k-AuNP with increasing \ch{K2CO3} concentrations. Solid lines of $R/R_F$ data in (a) and (d) are acquired from the best-fit electron density profiles in (b) and (e), respectively. 2D GISAXS data for (c) PNIPAM3k-AuNP and (f) PNIPAM6k-AuNP with increasing \ch{K2CO3} concentrations. Arrows in (c) and (f) indicate the calculated $Q_{xy}$ positions for a 2D hexagonal lattice.
 	}
 	\label{2d_3k6k_main}
 \end{figure*} 

Figure \ref{2d_3k6k_main} shows X-ray reflectivity (XRR) and grazing-incidence small-angle X-ray scattering (GISAXS) data from thin films of PNIPAM3k and PNIPAM6k grafted AuNPs at room temperature with various salt concentrations as indicated. The analysis of XRR in terms of ED profiles is shown in Figure \ref{2d_3k6k_main} (b) and (e). The broad bell-shaped peaks in the ED profiles are due to the AuNP cores. The diffraction pattern in Figure \ref{2d_3k6k_main} (c) and (f) show broad peaks corresponding to 2D short-range ordering where peak positions at $Q$ values are similar to that are in bulk. 

\clearpage
\subsubsection{Calculation of grafting density}
To determine the grafting density of PNIPAM on the AuNPs, we consider the scenario where PNIPAM3k-AuNPs have completely collapsed, releasing any trapped water. In this state, the overall diameter of the combined nanoparticle closely aligns with $d_{\rm NN}$. By measuring the difference in volume ($\Delta V$) between the bare AuNP ($D$ calculated from form factor measurements) and this combined structure, we can deduce the volume occupied by the grafted PNIPAM. This, in turn, can be employed to compute the weight of the attached PNIPAM ($W_{\rm PNIPAM}$) and, subsequently, the number of grafted chains per AuNP.

\begin{equation}\label{Eq:delv}
    \Delta V = \frac{\pi}{6} (d_{\rm NN}^3 - D^3) 
\end{equation}

Weight of PNIPAM:

\begin{equation}\label{Eq:weight}
    W_{\rm PNIPAM} = \Delta V \times \rho_{\rm PNIPAM}
\end{equation}

\begin{equation}\label{Eq:mass}
    \text{Number of chains} = \frac{W_{\rm PNIPAM}}{MW_{\rm PNIPAM}}
\end{equation}
Where $MW$ is the molecular weight of one PNIPAM chain.

Calculation for PNIPAM3k-AuNPs:
$d_{\rm NN}$ = 9.68 nm, $D$ = 8.7 nm , $\rho_{\rm PNIPAM}$= 1.1 $\frac{g}{cm^3}$ $MW$ = 4.9816 $\times$ 10$^{-21}$ g
Substituting these values to Equations \ref{Eq:delv},  \ref{Eq:weight}, \ref{Eq:mass}, we can estimate the number of chains grafted approximately equal to 28.78. 

Calculation for PNIPAM6k-AuNPs:
$d_{\rm NN}$ = 16.75 nm, $D$ = 8.6 nm , $\rho_{\rm PNIPAM}$= 1.1 $\frac{g}{cm^3}$ $MW$ = 9.9632 $\times$ 10$^{-21}$ g
Substituting these values to Equations \ref{Eq:delv},  \ref{Eq:weight}, \ref{Eq:mass}, we can estimate the number of chains grafted approximately equal to 235.15. Which is higher in number, indicating that it is not in a completely collapsed state with absorbed water.

\subsubsection{Calculation of nanoparticle molarity}
The molarity of AuNP suspensions is determined using the concentration data provided by the vendor (TedPella Inc.), which specifies the number of particles in a milliliter of solution. This calculation employs the following equation.

\begin{equation}\label{Eq:molarity}
    \text{Molarity (M)} = \frac{\rm Number~of~particles~per~ml}{N_{\rm A} \times 10^{-3} \rm ~liters}
\end{equation}
Where $N_{\rm A}$ is Avogadro's number.

\noindent For example, the molarity of 10 nm-sized bare AuNPs is calculated from the particle concentration provided on the vendor technical data sheet (in this case 5.7 $ \times 10^{12}$ particles/ml). Using equation \ref{Eq:molarity} yields of 9.46 $ \times 10^{-9}$ M, equivalent to 9.46 nM. Upon determining the molarity of the bare AuNPs, UV-visible spectrophotometry is used to ascertain their absorbance at 525 nm wavelength. Assuming the grafting does not affect the signal and aggregation does not occur, by comparing the absorbance profiles of the PNIPAM grafted AuNPs with those of the bare AuNPs, the concentration of the grafted AuNPs in suspension can be determined.

\end{document}